\documentclass[fleqn,usenatbib]{mnras}

\usepackage{newtxtext,newtxmath}

\usepackage[T1]{fontenc}

\DeclareRobustCommand{\VAN}[3]{#2}
\let\VANthebibliography\thebibliography
\def\thebibliography{\DeclareRobustCommand{\VAN}[3]{##3}\VANthebibliography}

\usepackage{graphicx}	
\usepackage{amsmath}	

\usepackage{tensor}
\usepackage{orcidlink}

\usepackage[utf8]{inputenc}
\usepackage{pgfplots}
\DeclareUnicodeCharacter{2212}{−}
\usepgfplotslibrary{groupplots,dateplot}
\usetikzlibrary{patterns,shapes.arrows}
\pgfplotsset{compat=newest}
\usepgfplotslibrary{external}
\newcommand{\einsteintoolkit}{\textsc{Einstein Toolkit}}
\newcommand{\thorn}[1]{\texttt{#1}} 
\newcommand{\vide}{\textsc{vide}}
\newcommand{\zobov}{\textsc{zobov}}
\newcommand{\gevolution}{\textit{gevolution}}
\newcommand{\LCDM}{$\Lambda$CDM}
\newcommand{\threericci}{\tensor[]{\mathcal{R}}{}}
\newcommand{\Hall}{\bar{H}}
\newcommand{\OmegaCurvature}{\Omega_\mathrm{K}}
\newcommand{\OK}{\Omega_\mathrm{K}}
\newcommand{\bOK}{\Omega_\mathrm{K}}
\newcommand{\OM}{\Omega_\mathrm{M}}

\newcommand{\OR}{\Omega_\mathrm{R}}
\newcommand{\OQ}{\Omega_\mathcal{Q}}

\newcommand{\HD}{H_\mathcal{D}}
\newcommand{\reff}{R_\mathrm{eff}}
\newcommand{\scaleradius}{R_\mathrm{s}}
\newcommand{\HSWradius}{R_\mathrm{v}}
\newcommand{\perhMpcUnit}{$h^{-1}$\,Mpc}
\newcommand{\perhMpc}[1]{#1\,\perhMpcUnit{}}
\newcommand{\AP}{Alcock--Paczy\'nski}
\defcitealias{Sachs1967}{Sachs--Wolfe (1967)}
\defcitealias{Rees1968}{Rees--Sciama (1968)}
\defcitealias{Buchert1997}{Buchert--Ehlers (1997)}
\defcitealias{Alcock1979}{\AP{} (1979)}
\defcitealias{Buchert2020}{BMR}
\defcitealias{wiltshireAverageObservationalQuantities2009}{TS}
\newcommand{\BMR}{\citepalias{Buchert2020}}
\newcommand{\BMRt}{\citetalias{Buchert2020}}

\definecolor{outerRegionColour}{rgb}{0.65,0.8,0.95}
\definecolor{innerRegionColour}{rgb}{0.35,0.45,0.95}
\definecolor{histColour}{rgb}{0.45,0.67,0.95}

\definecolor{outerRegionColourB}{rgb}{1.0,0.77,0.66}
\definecolor{innerRegionColourB}{rgb}{1.0,0.42,0.22}
\definecolor{histColourB}{rgb}{1.0,0.58,0.40}

\title[Void statistics in numerical relativity]{First investigation of void statistics in numerical relativity simulations}

\author[M. J. Williams et al.]{
Michael J. Williams,$^{1}$\thanks{E-mail: michael.williams@pg.canterbury.ac.nz}\orcidlink{0009-0000-2248-1931}
Hayley J. Macpherson,$^{2,3}$\orcidlink{0000-0002-9950-422X}
David L. Wiltshire,$^{1}$\orcidlink{0000-0003-1992-6682}
and Chris Stevens$^{4}$\orcidlink{0000-0002-0614-4879}
\\
$^{1}$School of Physical \& Chemical Sciences, University of Canterbury,
Private Bag 4800, Christchurch 8140, New Zealand\\
$^{2}${Kavli Institute for Cosmological Physics, The University of Chicago, 5640 South Ellis Avenue, Chicago, Illinois 60637, USA}\\
$^{3}${NASA Einstein Fellow}\\
$^{4}$School of Mathematics \& Statistics, University of Canterbury,
Private Bag 4800, Christchurch 8140, New Zealand
}

\date{Accepted 2024 December 9. Received 2024 November 25; in original form 2024 March 28}

\pubyear{2024}

\begin{document}
\label{firstpage}
\pagerange{\pageref{firstpage}--\pageref{lastpage}}
\maketitle

\begin{abstract}
We apply and extend standard tools for void statistics to
cosmological simulations that solve Einstein's equations with numerical relativity (NR).
We obtain a simulated void catalogue without Newtonian approximations,
using a new watershed void finder 
which operates on fluid-based NR simulations produced with the \einsteintoolkit{}.
We compare and contrast measures of void size and void fraction,
and compare radial stacked density profiles to empirically-derived
Hamaus--Sutter--Wandelt (HSW) density profiles and profiles based on distance to void boundaries.
We recover statistics 
roughly consistent with Newtonian $N$-body simulations where such a comparison is meaningful.
We study variation of dynamical spatial curvature and 
local expansion
explicitly demonstrating the spatial fluctuations of these quantities in void regions.
We find that voids in our simulations expand $\sim$10--30\% faster than the global average and the spatial curvature density parameter in the centre of voids reaches $\sim$60--80\%.

\end{abstract}



\section{Introduction}

A fundamental aspect of the Universe is the organisation of matter into clusters and filaments, forming the large-scale structure that is central to modern cosmology.
The clustering of matter into dense structures leads to the emergence of cosmic voids,
which make up most 
of the volume of the Universe
and constitute its largest typical observed structures
\citep{Hoyle2002, Hoyle2004, Pan_2012}.

Observations present us with the immediate dichotomy that voids -- the absence of matter -- by their very nature can only be understood by directly observing the complex matter distribution within which they grow and interact. Observations need to be combined with modeling and numerical simulations to glean any understanding. The goal of this paper is to use numerical relativity to extend that modeling to a full general relativistic analysis of void statistics for the first time.

These aims require not only that we take care with approximations derived for the standard $\Lambda$ Cold Dark Matter ($\Lambda$CDM) cosmology, but also those for the generic class of Friedmann--Lema\^{\i}tre--Robertson--Walker (FLRW) models on which $\Lambda$CDM is based. We study realistic simulations for which the FLRW approximation is extremely good in the early universe, but which at later epochs admit the non-linear features intrinsic to general relativity (GR).

Cosmic voids have uses for studying a wide range of cosmological properties and effects \citep{Moresco2022}.
Their dynamics remain near-linear even in the present epoch, 
\citep{Stopyra2020, Schuster2023}, allowing for the simpler extraction of cosmological information.
Observed or simulated void statistics can be used to constrain cosmological parameters
\citep{Lavaux2010, Bos2012, Hamaus2016, Contarini2023}
and can provide another perspective on tensions in the standard model of cosmology
\citep{Contarini2024}.

Simulated void catalogues are typically produced by Newtonian $N$-body simulations, in which 
expansion is prescribed by scaling the simulation volume
according to the Friedmann equation from an FLRW model.
However, in GR
matter and space-time curvature are non-linearly related.
Consequently, regional expansion in the universe is determined by its local matter content, allowing over-dense and under-dense regions to have varying degrees of spatial curvature
in any foliation. 
This may occur
even if the average spatial curvature of the universe is zero (as in \LCDM{}). 
This is particularly relevant
for cosmic voids,
which have a negative spatial curvature
and expand at a greater rate than the over-dense sheets, filaments and knots of the cosmic web
\citep{Peebles_1993,Peebles_2001,Macpherson2018}. 
These concepts exist in standard cosmology. However, they are not explicitly implemented in the treatment of space-time in Newtonian simulations. In principle, variations of curvature and expansion in voids could be calculated from the Newtonian potential, but such an analysis has not yet been done.

A notable exception to traditional $N$-body methods is the heuristic {\sc AvERA} scheme \citep{Racz2017} which replaces the Friedmann equation by an evolution code step that applies a volume-average to the scale factor governing the $N$-body simulation volume. This scheme has had phenomenological success in application to tensions in the integrated Sachs--Wolfe effect \citep{Beck_2018,Kovacs_2020,Kovacs_2022}, raising the question of whether such results can be replicated in numerical relativity.

Traditional $N$-body codes have been been significantly improved, most notably by {\gevolution} \citep{Adamek2016} which uses the weak-field limit of GR. 
In simplified setups, {\gevolution} has been shown to agree with 
numerical relativity 
\citep{Adamek2020,Adamek2024}. 
However, the weak-field scheme inherently precludes a full investigation of the potential for non-linear GR effects to \textit{significantly} alter the chosen global background\footnote{While a particular background cosmology must be chosen to set initial data in \textit{gevolution}, in principle this can change during the evolution due to the absorption of homogeneous modes of the perturbations into the scale factor. However, this change in background is restricted to be small \citep[see Section~5.3 of][]{Adamek2016}.}.

Another alternative to evolving all degrees of freedom contained in the Einstein equations is to self-consistently remove degrees of freedom with likely small contributions in cosmology: 
{\bf(i)} gravitational waves in the {\sc GRAMSES} \citep{gramses1, gramses2} code; 
{\bf(ii)} vorticity and the magnetic Weyl curvature in quiet universes \citep{Heinesen2022};
and {\bf(iii)} all of the above in silent universes \citep{Bolejko2018silent}. 
Silent universe models have been used to seek solution to the Hubble tension \citep{Bolejko2018Hubble}.
However, the relationship between the \citet{Bolejko2018Hubble} solution and realistic initial conditions from the cosmic microwave background (CMB) epoch are not clear. 
In this work, we are interested in using numerical relativity to circumvent such limitations and study the fully non-linear problem. 

\subsection{Numerical relativity}\label{sec:NR}
Numerical relativity (NR) was developed for the study of compact objects, such as binary black holes \citep[e.g.,][]{Pretorius2005,Campanelli:2006,Baker:2006} and neutron stars \citep[e.g.,][]{Shibata:2005,Baiotti:2008}. It has since been used for a variety of applications across relativistic astrophysics \citep[see][for a review]{Lehner:2014} and early-Universe cosmology \citep[e.g.,][]{Clough:2017,Giblin:2019pre}.

NR has been adopted to study late-time inhomogeneous cosmology in the absence of space-time symmetries,
beginning with \citet{Giblin2016} 
and the \textsc{CosmoGRaPH} code \citep{Mertens2016}, 
and simultaneously \citet{Bentivegna2016} using the \einsteintoolkit{},
independently followed by \citet{Macpherson2017},
also using the \einsteintoolkit{} \citep[see also][]{Daverio2017,Wang2018,Tian:2021}.
These groups adopt a continuous fluid approximation for matter,
rather than an $N$-body particle ensemble.
More recently, \citet{Giblin2019}, \citet{Daverio2019} and \citet{East2019} have performed NR simulations coupled to $N$-body particle dynamics 
\citep[see also][for coupling to smoothed particle hydrodynamics (SPH)]{Magnall2023}.
In this work, we will use the \einsteintoolkit{}, which adopts a fluid approximation for the matter content. 
This approximation breaks down below the scale of the largest bound structures, namely, galaxy clusters.
For this reason we will ensure our simulations only sample scales on which such an approximation should be valid.

NR allows for the description of spacetime
without the assumption of closeness to any background metric.
We solve Einstein's equations in full,
with no linearisations or other approximations of the metric,
allowing us to capture non-linear phenomena that would be missed by such assumptions.
With access to the full metric throughout the simulation,
we are able to calculate quantities such as 
the local expansion rate, the Ricci scalar,
and the kinetic curvature parameter,
which depend on the local values of the metric components. 

\subsection{Backreaction of inhomogeneities}

It is conventionally assumed that departures from average isotropic cosmic expansion can always be exactly reduced to local Lorentz boosts -- i.e., peculiar velocities of the source, observer and intervening structures -- on a global FLRW background. However, in GR the growth of structure may lead to an expansion history significantly different to any single global FLRW background. Such cosmological backreaction scenarios \citep{Buchert2000,Buchert:2012A,Buchert2020} may in turn affect the traditional interpretation of peculiar velocities.

Different types of backreaction have been classified in the well-known \citet{Buchert2020} \BMR{} formalism,
which provides a set of equations for the evolution of scalar averages of small scale structures.
The first Buchert equation may be written as the
{\em energy density sum-rule}:\footnote{
A cosmological constant backreaction density, $\Omega_\Lambda$, and additional backreaction terms are also found in general \BMR{}, but are not relevant for interpreting our simulations. The radiation density $\OR$ is not relevant except insofar as the ratio $\OR/\OM\simeq0.3$ at last scattering, with significance for calibrations relative to CMB data. Our notation differs from that of \BMRt{}, who use $\Omega_\mathcal{R}$ in place of $\bOK$, while not explicitly writing a radiation density. Positive values $\bOK>0$ correspond to 
negative spatial curvature, as is the case for void domination.} 
\begin{equation}
\label{eq:sumrule}
\OM+\OR+\bOK+\OQ=1\,.
\end{equation}
Here $\OM$ and $\OR$ are analogous to matter and radiation energy densities
in the Friedmann equation, but scales with respect to powers of an 
{\em average volume scale} $a_\mathcal{D}$,
(cf.\ \eqref{eq:scalefactor} below), 
{\em not} a background metric scale factor; $\bOK$ is the average spatial curvature or {\em kinetic curvature} density; and $\OQ$ is the {\it kinematical backreaction} density.

Much attention has focused on the magnitude of 
$|\OQ|$, going back to the 2000s \citep{Ishibashi2006}. However, while a nonzero $|\OQ|$
is necessary for realistic cosmological backreaction,\footnote{
The imposition of periodic boundary condition in Newtonian cosmologies -- the so-called {\em torus condition} -- results in $\OQ$ 
necessarily vanishing in averages over the full global spatial hypersurfaces. The heuristic {\sc AvERA} scheme \citep{Racz2017} explicitly breaks the assumptions of the \citetalias{Buchert1997} theorem, and was criticised \citep{Kaiser2017,Buchert2018} as a result. However, the quasilocal nature of gravitational mass in GR means that laws of mass conservation in GR may differ from their Newtonian counterparts. Thus the {\sc AvERA} phenomenology may point to deeper fundamental questions.}
it may be small, 
as in the case of the timescape model \citep{wiltshireCosmicClocks2007a,wiltshireExactSolution2007b,wiltshireAverageObservationalQuantities2009, DuleyTimescapeRadiation2013}.
By contrast, a positive kinetic curvature term $\bOK$ makes a very significant contribution to the late epoch energy budget \eqref{eq:sumrule} in several backreaction models, including those with an appreciable cosmological constant \citep{Bolejko2018Hubble}.
In the timescape 
model, 
$\bOK$ has a direct physical interpretation as the quasilocal (i.e., regional) kinetic energy of expansion of voids --
producing an \emph{apparent} acceleration that does away with dark energy.

Inhomogeneities are the source of local and quasilocal anisotropies in both GR and its weak field Newtonian limit.
In fully relativistic cosmology regional anisotropies are not limited to a global FLRW background plus the spherical multipole expansion obtained by local Lorentz boosts in a $v/c$ power series. The differences have been classified via multipole moments as {\em nonkinematic differential expansion} \citep{Bolejko2016,McKay2016,Damthesis2016}.

The methods of Section \ref{sec:NR} have enabled the first numerical investigations into the impact of backreaction 
\citep{Bentivegna2016,Macpherson2018,Macpherson2019} 
and 
cosmological observables 
\citep[e.g.][]{Giblin2017,Macpherson2023} in fully non-linear late-time cosmology. 
The characteristic features of quasilocal anisotropies expected in GR, as constrained by the initial matter power spectrum, are an important fundamental question.
This paper quantifies the principal contribution to these small scale anisotropies -- cosmic voids -- for the first time.

\subsection{Outline}

Section \ref{sec:simulations} details the NR simulations which produce the void sample, 
and discusses the advantages and drawbacks compared to the typical use of Newtonian $N$-body simulations.
Section \ref{sec:voidfinder} describes the operation of watershed void finders in general,
the specifics of our implementation,
and the differences from other watershed void finders that are necessitated by our simulations.
Section \ref{sec:statistics} covers void statistics in the simulations.
As well as standard statistics such as the void size function and radial density profiles,
we also produce radial profiles of three scalars derived from metric.
We also use the technique of \citet{Cautun2016}
to produce profiles based on distance to void boundaries.
In Appendix~\ref{a:resolution} we show that simulation numerical error is not large enough to affect our statistics.

\section{Numerical Relativity Simulations}
\label{sec:simulations}

We perform analysis of void statistics on a pair of cosmological simulations
produced using the \einsteintoolkit{}\footnote{\url{https://einsteintoolkit.org}} \citep{EinsteinToolkit:2023_05},
an open-source NR platform of computational tools for gravitational physics and relativistic astrophysics.

Our simulations are initialised at redshift $z \approx 1000$ 
using \thorn{FLRWSolver} \citep{Macpherson2017},
which generates linear\footnote{A very good approximation to the full non-linear constraints for the very small perturbations at the initial redshift.}
fluctuations about an Einstein de-Sitter (EdS) background space-time. The perturbations are drawn from a matter power spectrum
output from CLASS \citep{CLASS}.
Importantly, while
our initial data utilises an FLRW background,
no such background is explicitly assumed during evolution.

We evolve the simulations from the initial redshift through to $z\approx0$ 
using the 3+1 Baumgarte--Shapiro--Shibata--Nakamura--Oohara--Kojima \citep*[BSSNOK or BSSN,][]{Baumgarte1998, Shibata1995, Nakamura1987} formalism of NR. 
The final slice of the simulation is defined based on the predicted change in volume from the initial background EdS model. The space-time is evolved using the \thorn{McLachlan} \citep{Brown:2009McLachlan} \thorn{ML\_BSSN} thorn in the ET and the hydrodynamics is evolved using \thorn{GRHydro} \citep{Baiotti:2005GRHydro}. 
We approximate matter as dust, namely, 
as a pressureless\footnote{\thorn{GRHydro} does not allow for an exactly zero pressure, so we choose $P\ll\rho$ using a polytropic equation of state (EOS), which is sufficient to approximate dust \citep{Macpherson2017}. Such an EOS has also been shown to be beneficial in analytic studies in inhomogeneous cosmology \citep{BolejkoLasky2008}.} perfect fluid (under a continuous fluid approximation). 
We set the shift $\beta^i=0$ throughout and choose a harmonic-type slicing to evolve the lapse $\alpha$ according to \citep[as in][]{Macpherson2019} 
$\partial_t \alpha = -1/3 \, \alpha^2 K$ where $K$ is the trace of the extrinsic curvature. In the FLRW limit -- or in the case of linear perturbations -- this gauge coincides with a conformal-time slicing. 
When calculating the expansion and curvature for our void profiles, we calculate quantities intrinsic to the fluid flow in the simulation. These quantities (defined in Section~\ref{sec:curvatureProfiles}) represent qualities of the frame that is everywhere orthogonal to the fluid 4-velocity, $u^\mu$.
This fluid rest-frame
is a more physically-motivated spatial slice than the particular slices we choose for the simulation itself. 
See \citet{Macpherson2017} and \citet{Macpherson2019} for further technical specifics of the simulation set up.

In this work, we analyse simulations with two different physical resolutions:
grid cell side lengths of
\perhMpc{4} and
\perhMpc{12}.
These scales are the smallest we can reliably simulate under the assumption of a continuous dust fluid, since below these scales both velocity dispersions and baryonic effects become important.
Both simulations have a grid resolution of $256^3$ cells, 
yielding a side length of \perhMpc{1024} and \perhMpc{3072}, respectively, at the final redshift $z \approx 0$.

When generating the initial conditions, we
remove power at all wavelengths below 
$\sim 10$ grid cells from the power spectrum. Specifically, we set $P(k>k_{\rm cut})=0$ where $k_{\rm cut}=2\pi/\lambda_{\rm cut}$ and $\lambda_{\rm cut}\approx 10\Delta x$ (where $\Delta x$ is the grid cell size). The power at $k\leq k_{\rm cut}$ is set according to the input power spectrum from CLASS. 
Removing these modes reduces the numerical error associated with under-sampling small-scale structures \citep[see also][]{Giblin2017}. 
The different physical resolutions between the two simulations means that the initial smoothing scales are \perhMpc{40} and \perhMpc{120}, however, structure can (and does) form below this scale at late times in the simulation due to non-linearity (the amplitude of this structure is significantly damped).

Our simulations do not include a cosmological constant, $\Lambda$, in the evolution of Einstein's equations. The ET has not yet been benchmarked for use with 
a cosmological constant, and this is the focus of ongoing work.
Including a cosmological constant would be expected to alter the void statistics,
giving larger voids due to their accelerated growth in the most recent epoch of expansion history.

\section{Watershed Void Finder}
\label{sec:voidfinder}

We present a new watershed void finder,
the \textsc{Watershed Einstein Topography} (WET) void finder,
designed for use in fluid-based NR
simulations.
Watershed void finding is now the most commonly used means of extracting catalogues of voids from data, described in detail later in this section.
Conceptually, the watershed transform partitions the density field into segments, or ``basins'',
where each segment contains all the points that can follow a descending path (i.e., decreasing density) to a particular local minimum.
Note that by construction, there is exactly one watershed segment per local minimum.

This process can be understood by analogy to
rainwater moving down a mountainous landscape,
where water moving from
high elevation collects at a local minimum at lower altitude, 
while ridges in the terrain 
separate these basins.
Applied to cosmology, instead of landscape height we have the matter density, so this process
identifies density depressions in the matter field, 
some of which we might define as cosmic voids.
Sheets, filaments, and nodes form the visible aspects of the large-scale structure in the Universe, 
and these are the ridges in the density field that separate voids.

This method originates with the Watershed Void Finder \citep*{Platen_2007}
and \zobov{} \citep{Neyrinck2008}.
One popular watershed void finder based on \zobov{} is \vide{} \citep{sutter2014vide}, 
which has been used 
in the study of both observational data
\citep[e.g.,][]{Sutter2012, Sutter2014AP}
and simulations \citep[e.g.,][]{Contarini2022}.

Most void-finding software is designed to operate on sets of particle positions, 
either from observational data (positions of galaxies)
or
the output of $N$-body simulations \citep{Platen_2007, Neyrinck2008, sutter2014vide}.
For watershed void-finding,
it is necessary to translate these into a continuous density field
upon which the watershed transform can be performed. 
It is typical
\citep[e.g., \zobov{};][]{Neyrinck2008}
to construct this continuous density field through Voronoi tessellation
\citep[or like][use its dual, the Delaunay tessellation]{Platen_2007},
where each galaxy or simulation particle is assigned all of the volume that is closer to it than to any other particle.
The density for that whole region is then determined by dividing the particle's mass by the volume, as if the matter had been spread evenly around the whole region.

As our simulations are fluid-based, 
we have direct access to all variables at every point on the grid.
Thus, we do not need to perform a density field estimation in the above manner.
However, in fluid-based simulations the density field cannot be as finely resolved as
compared to high-density regions of particle-based data,
such as in the structures of the cosmic web.
However, we do not have the issue of shot noise in the estimation of density when particles are sparse,
such as near void centres,
though that is less of an issue for the statistics we look at here, which involve averaging over many voids.

\subsection{Watershed segmentation}
As a density field estimation from particles is unnecessary in our case, our first step is the watershed transform itself,
to segment the grid into watersheds. 
Here we give a short description of the algorithm used 
by the WET void finder.
For a detailed description of the watershed void-finding process, see \citet{Platen_2007}.
 
Beginning with the least dense cell, 
we iterate through each cell in the simulation volume
in increasing order of density. 
Each cell is to be assigned an index representing which watershed segment it is part of.
If the cell neighbours only one segment, then it is added to that segment.
If none of a cell's neighbours are yet part of a segment,
the cell must be a local minimum, so a new segment is defined.
Cells that neighbour more than one distinct segment are not added to any segments,
and instead are marked as segment boundary cells, which make up the density ridges separating segments.
These ridges are related to the sheets and filaments of galaxy clusters that separate cosmic voids.

\begin{figure}
    \centering
	\includegraphics{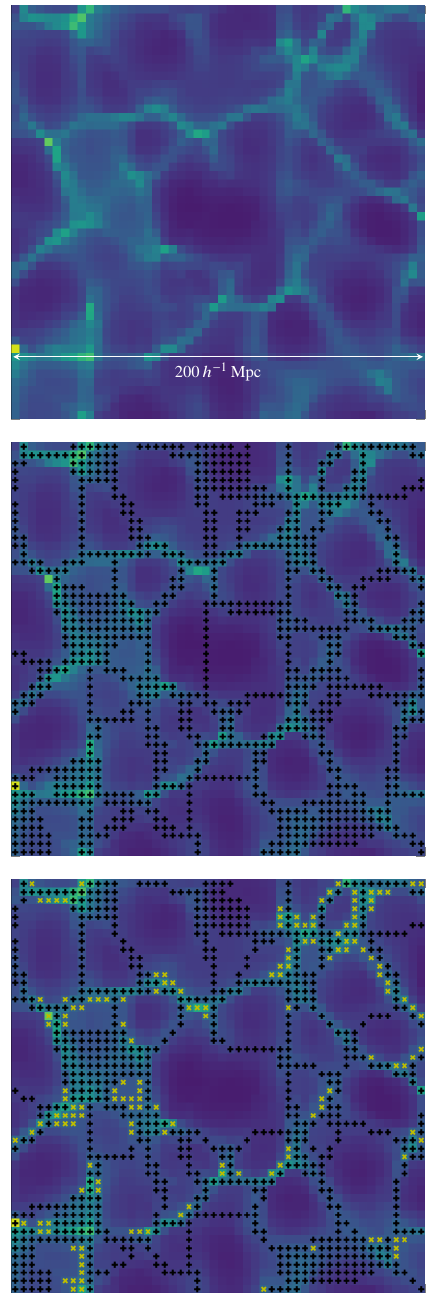}
    \caption{Density on a (\perhMpc{200})$^2$ region of a slice through the \perhMpc{4} resolution simulation. 
    Brighter green, towards yellow, indicates high density, and darker purple areas are of low density. 
    The top panel shows the raw density field, 
    the centre panel is annotated showing the segment boundary cells (black +), and the lower plot is annotated showing the void wall cells (black +) and over-dense excluded cells (yellow \textsf{x}).}
    \label{fig:voidSlice}
\end{figure}

Fig.~\ref{fig:voidSlice} shows a demonstration of the algorithm.
The top panel shows the density on a \perhMpc{200} region of a two-dimensional slice through the \perhMpc{4} resolution simulation.
The centre panel
shows the same slice after the above segmentation process has been applied to the whole three-dimensional volume,
with the segment boundary cells marked with black crosses.
The contiguous areas of marked cells correspond to boundary walls that lie in the plane of the slice,
separating a segment above the slice from a segment below.
The segments produced mostly correspond to the voids in the density field which can be seen by eye.
However, not every density ridge necessarily corresponds to significant physical structure separating two voids,
so further steps are necessary to translate the identified segments and boundaries into voids and their walls.

\subsection{Merging segments}

Large voids are known to have a rich substructure \citep{Tikhonov2006, Sutter2014VoidLife}; containing smaller voids and low-density structures.
This can result in a
slight density ridge inside a void,
and the segmentation process will identify the areas on each side of such a ridge as separate watershed segments. 
However, it may be reasonable to consider both segments part of the same void.
To this end, we perform a merging step after segmentation,
which identifies segments that are separated by a wall with density below a chosen threshold and 
combines them into a single larger void.
Any ridge with a density contrast below this threshold will be considered as part of the larger void surrounding it.

We use the density contrast, $\delta$, to define this threshold density: 
\begin{equation}
    \delta \equiv \frac{\rho}{\bar\rho} - 1\,,
\end{equation}
where $\rho$ is the rest-mass density and
$\bar\rho$ is its average over the spatial simulation slice.
We choose $\delta=-0.8$
as our threshold,
motivated by the density at which a top-hat model of an expanding under-density in an 
EdS universe experiences shell crossing \citep{Sheth_2004}.
Further, this same threshold and merging process is used in widely-used watershed void finders.
It is a default for \zobov{} \citep{Neyrinck2008},
a commonly used option in \vide{} \citep{sutter2014vide},
and has been used in other implementations \citep[e.g.,][]{Cautun2016}.

The lower panel of Fig.~\ref{fig:voidSlice} shows the results of merging the segments found in the previous step (centre panel).
Now, the  black `+' markers indicate void boundaries.
As a result of this step, the large void in the centre of the figure
is no longer split into two segments by a wall 
(as in the centre panel),
as the density ridge is below our threshold for significance. 

\subsection{Density-based exclusion}
\label{sec:exclusion}
By construction, the barriers between segments are only one grid cell thick,
but the over-dense walls separating physical voids may be thicker in practice. 
To account for this, we remove any cell from its void if it has a density greater than the whole-volume spatially-averaged density.
Thus, no over-dense cell can ever be part of a void, and are always considered walls,
but under-dense cells may be either part of voids or walls. 
Additionally, a watershed segment can in principle have any minimum density,
so any slight depression in the density field, even in an over-dense area, will be considered a watershed segment.
This threshold prevents such depressions from being considered cosmic voids in our analysis.

The lower panel of Fig.~\ref{fig:voidSlice} shows this exclusion process, with the excluded cells annotated with a yellow `\textsf{x}' marker.
The necessity of this step can be seen, for example, 
at the top of the large central void, where an over-dense region was
previously included in the segment
but has now been excluded.

\subsection{Volume calculation}
We calculate the volume of a void 
by summing the volume of the grid cells contained within it.
The volume of a cell is defined as 
\begin{equation}
\label{eq:cell_volume}
    V_\mathrm{cell} = \sqrt{\gamma}\,\Delta x^3,
\end{equation}
where $\gamma$ is the determinant of the spatial metric in the grid cell
and $\Delta x$ is the physical resolution \perhMpc{4} or \perhMpc{12}.

The volume element $\sqrt{\gamma}$ implies that not all cells have 
the same volume, a quality unique to our general-relativistic treatment.
For the simulations we use here, we find variations in volume which are always <0.13\%. 
Such a small variance in volume is expected in cases where the spatial metric remains close to a background FLRW expansion (with a metric perturbation $\phi\ll 1$), as is the case in our simulations \citep[see][]{Macpherson2019}.
However, the ability to account for this effect in 
our void finder 
could be important in cases where the metric perturbations are larger.

Some of the features identified as 
voids are very small, 
only one or two grid cells in radius.
We discard these voids since they  
are too close to the grid scale to be resolved.
This excludes approximately 7\% of potential voids in the \perhMpc{4} resolution simulation, and approximately 8\% at \perhMpc{12} resolution.

\subsection{Differences compared to particle-based methods}

Void-finding on particle-based data
must account for shot noise when reconstructing the density field (though Voronoi tessellation suffers much less from this than other methods),
which can produce apparent depressions or walls which are not associated with actual structures.
Fluid-based simulations do not have such issues,
so our void finder skips several corrective steps common to particle-based void finders that are unnecessary for our data.

The walls that our void finder finds to separate voids must be a minimum of one grid cell thick.
Physically, at our two resolutions, this means a thickness of \perhMpc{4}  
and \perhMpc{12}. 
If finer detail could be resolved, it is likely that
the wall is thinner than this, 
and therefore that the void should be
larger than the volume we identify.
This is particularly important for simulations with lower physical resolution.
This is also expected to impact our void density profiles,
as the matter is spread out over an entire grid cell.
Since $N$-body simulations are capable of resolving smaller scales
(particularly in walls, where the number density of particles is high), 
these will identify a more realistic wall structure.

\section{Void Statistics}
\label{sec:statistics}
In this section, we outline the statistics of the sizes and shapes of voids in our simulations.
We find a total of 30519 voids above the size cut-off 
in the (\perhMpc{1024})$^3$ volume of the \perhMpc{4} resolution simulation,
and 39206 in the (\perhMpc{3072})$^3$ volume at \perhMpc{12} resolution.

\subsection{Void size function}
\label{sec:vsf}
The void size function describes the number density of voids of different sizes.
This statistic has been modelled analytically \citep[][]{Sheth_2004}
and is useful for constraining cosmological parameters, in particular
for the study of dark energy \citep{Contarini2022}.
As voids are not perfectly spherically symmetric,
it is common to define the effective radius
of a void as the radius of a sphere with the same volume as the void, namely,
\begin{equation}
\label{eq:reff}
    \reff = \left(\frac{3}{4\pi} \sum_\mathrm{cells} V_\mathrm{cell} \right)^{\frac{1}{3}},
\end{equation}

\begin{figure}
	\includegraphics{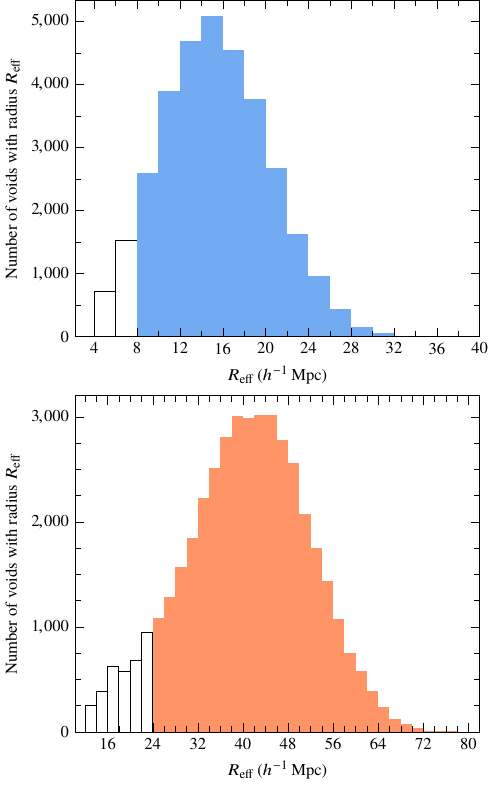}
    \caption{
    Void size function for the simulation at a resolution of \perhMpc{4} (top panel) and \perhMpc{12} (bottom panel), 
    binned by effective radius.
    The un-shaded bars show radii where voids are excluded due to low resolution.}
    \label{fig:vsf}
\end{figure}

Fig.~\ref{fig:vsf} shows the void size function for the \perhMpc{12} resolution simulation (top panel) and the \perhMpc{4} resolution simulation (bottom panel).
The largest void in the \perhMpc{4} simulation has an effective radius of $\reff\approx$ \perhMpc{40},
while the largest in the \perhMpc{12} simulation have $\reff\approx$ \perhMpc{80}.

The void size function depends on the smoothing scale 
\citep{Sheth_2004}, evident from the differences between the two panels in Fig.~\ref{fig:vsf}, 
so different simulations and observational datasets will also differ in their void size functions
because of their various smoothing scales \citep[see e.g.,][for a study showing vastly different void size functions\footnote{
Specifically, in the \textit{isolated} (no ``segment merging'') catalogue of \citet{Schuster2023}, comparable to our approach, VSFs for different resolutions differ, but their \textit{merged} catalogue containing all subvoids and parent voids shows agreement for abundance of large voids at different resolutions.} in different samples]{Nadathur2014, Schuster2023}.
Further, void size may be affected by the accelerated expansion resulting from a cosmological constant,
so it is difficult to make a quantitative comparison between the void size function in this simulation without a cosmological constant
and Newtonian $N$-body simulations that include a cosmological constant in the Friedmann equations that define the expansion of their Universe.

Qualitatively, the shape of our void size function is consistent with
past studies using $N$-body simulations and from observation,
showing a skewed bell-shaped distribution,
with a tail in the positive direction but
declining rapidly as radius decreases.

\subsection{Void fraction}

Voids comprise the majority of the volume of the late-time Universe. Constraining the precise fraction of voids is important for the timescape model of cosmic backreaction \citep{wiltshireCosmicClocks2007a,wiltshireExactSolution2007b,wiltshireAverageObservationalQuantities2009, DuleyTimescapeRadiation2013}, since it is the free parameter which replaces $\Omega_{\rm M}$ in standard FLRW models.

In the \perhMpc{4} resolution simulation, 
61.5\% of the volume is in cells marked as being part of a void,
and for the \perhMpc{12} resolution simulation this fraction is 50\%.
This value does not include density depressions that are excluded on the basis of their small size 
(see Section~\ref{sec:exclusion}).

The void volume fraction is also dependent on the definition of what it means for a region to be part of of a void,
for which there is no strict settled definition.
As such, different void-finding techniques can find different void fractions in the same data.
For example, \citet{Pan_2012} find a void fraction of 62\% in the Sloan Digital Sky Survey Data Release 7. 
\citet{Pan_2012} do not use a watershed void finder, 
but rather a nearest-neighbour method based on \citet{El_Ad1997}. 
The authors find voids with a minimum effective radius of 10 Mpc, median of 17 Mpc, and maximum of just over 30 Mpc.
\citet{Nadathur2014} run \zobov{} on the same survey, 
dividing it into different samples based on redshift bins.
They find void fractions within these samples between
30\% and 42\% under their basic definition of a void (and lower fractions for stricter definitions).
This illustrates the difficulty in producing a single well-defined value for the void fraction of a given dataset.

\subsection{Finding void centres}
\label{sec:centres}
In Section~\ref{sec:radialProfiles},
we construct radial profiles of the voids for 
density, expansion rate, the Ricci scalar, and the kinetic curvature parameter.
These profiles are usually constructed using the radial distance from the void center.
However, voids are not exactly spherical,
so there are several possible ways to define a point as the centre of the void from which to measure radial distance
\citep{Nadathur2015}.
For particle-based data -- such as Newtonian $N$-body simulations or galaxy surveys --
the void centre is typically defined as the volume-weighted barycentre 
\citep[e.g.,][]{Sutter2012,Nadathur2014},
given by
\begin{equation}
   \mathbf{X}^\mathrm{bc}_v = \frac{1}{\sum_i V_i} \sum_i  V_i\,\mathbf{x}_i\,, 
\end{equation}
where $\mathbf{x}_i$ is the position of the $i^{\mathrm{th}}$ 
particle in a void $v$,
and $V_i$ is the Voronoi cell volume associated with it.
In our case, with a uniform grid of cells of unequal volume,
this can be generalised to
\begin{equation}
   \mathbf{X}^\mathrm{bc}_v = \frac{1}{\sum_\mathrm{cells} \sqrt{\gamma}} \sum_\mathrm{cells}  \sqrt{\gamma}\, \mathbf{x}\,.
\end{equation}
This definition does not take into account the matter distribution inside the void (except indirectly by a very small amount, due to $\sqrt{\gamma}$), only the void's overall shape.
Extremely low density regions are weighted just as much as barely under-dense regions, provided that the latter are included in the void.
As a void is characterised by extreme under-density, this property may not be desirable, as we may wish for the centre to be closer to less dense regions.
In light of this, we also consider another means of finding the centre,
to take into account the internal matter distribution of the void.
This alternative measure of centre is related to the centre of mass,
but weighted towards areas of low density (i.e., the interior of the void) rather than high density (the walls), given by a weighting function $f(\rho)$.
\begin{equation}
   \mathbf{X}^\mathrm{cm}_v = \frac{1}{\sum_\mathrm{cells} f(\rho) \sqrt{\gamma}} \sum_\mathrm{cells} f(\rho) \sqrt{\gamma}\,\mathbf{x}\,. 
\end{equation}
We consider two choices of $f(\rho)$ which ensure that the centre is weighted towards areas of low density,
namely $f(\rho)=1/\rho$ and $f(\rho)=\rho_\mathrm{ave} - \rho$.
When matter is distributed unevenly around a void, its density-weighted centre will differ from its centre of volume.
Under these definitions, the decision to include or exclude a higher-density cell has less impact on the centre that it would have on the centre of volume. 
Thus, it may be more robust to the choice of exclusion threshold or slight changes in the density near it, which may cause the set of excluded cells to change.

In the \perhMpc{4} resolution simulation, 
the mean distance between a void's density-weighted centre $\mathbf{X}^\mathrm{cm}_v$ and its centre of volume $\mathbf{X}^\mathrm{bc}_v$
is approximately \perhMpc{0.78}, or ~0.20 grid cells, for $f(\rho)=1/\rho$, 
and \perhMpc{0.88} (0.22 grid cells) when $f(\rho)=\rho_\mathrm{ave} - \rho$.
In both cases the difference has an upper bound of a little over one cell.
The mean distance between the two definitions of $\mathbf{X}^\mathrm{cm}_v$ is \perhMpc{0.27}.
Since these differences are small with respect to our grid resolution, we consider all three to be appropriates measures of centre in our case. 
We use the inverse-density weighted centre $\mathbf{X}^\mathrm{cm}_v$, $f(\rho)=1/\rho$, hereafter 
for the sake of comparison with existing literature.

\subsection{Radial profiles}
\label{sec:radialProfiles}
\begin{figure*}
	\includegraphics{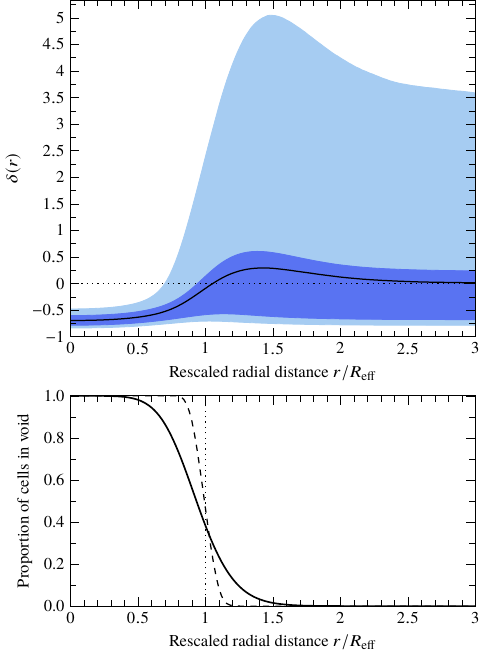}
	\includegraphics{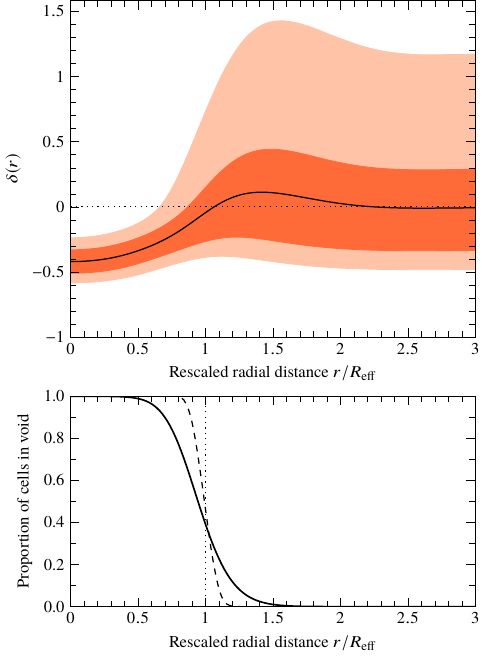}
    \caption{
    Upper: The 
    stacked density contrast (solid curve) on a spherical shell at distance $r$ from the stacked void centre in the \perhMpc{4} (left) and \perhMpc{12} (right) simulations,
    with each void individually normalised by its effective radius $\reff$. The shaded areas show the middle ~68.5\% and ~95\% of $\delta$ values for cells on that shell.
    Lower: The fraction of cells inside the void as a function of radius (solid) compared to perfect spheres (dotted) and a set of grid-approximated spheres with radius 3--5 grid cells (dashed).
    }
    \label{fig:radial_density}
\end{figure*}

As voids are often approximated as being spherical,
it is common to consider how the density varies radially from the centre.
\citet{Sheth_2004} suggest a theoretical radial density profile based on the excursion set formalism.
Voids in data are not
spherically symmetric,
so it is necessary to stack a large number of voids 
(i.e., align their centres and average over all voids, after rescaling by effective radius)
to form a structure that is spherically symmetric,
then to determine density as a function of distance from the stacked centres.
Such stacked averages have been shown to be accurately described by a universal radial density profile,
known as the HSW profile
\citep*{Hamaus2014},
\begin{equation}
\label{eq:hsw}
    \delta = \delta_c\frac{1-(r/\scaleradius)^\alpha}{1+(r/\reff)^\beta}\,.
\end{equation}
Here, $\delta_c$ is the void's central density contrast,
$\scaleradius$ is a scale radius at which $\delta$ crosses zero,
and $\alpha$ and $\beta$ are parameters that determine the inner and outer slopes of the wall, respectively.
This gives a four-parameter model for density contrast in stacked voids. 

\subsubsection{Constructing stacked radial profiles}

We construct the radial profile of an individual void
by
choosing 500 random directions (the same set of directions for each variable)
and then sampling the variable at intervals of $0.02\reff$ along each direction from the center, to a maximum distance of $3\reff$.
The average value of the variable over the spherical shell consisting of all directions' sample points at that radial step gives the corresponding value for the spherically averaged profile.

Because the steps are in terms of a fraction of each void's radius,
individual profiles can easily be stacked and averaged to produce a radial profile for a sample of voids.
The process of stacking a large number of voids produces a spherically symmetric object,
despite individual voids not being necessarily spherical.
Stacking voids in this manner has other uses beyond extracting radial profiles.
It can also be used to amplify the signal of the integrated Sachs--Wolfe effect \citep{Granett2008},
as temperature fluctuations in the CMB are larger than the temperature change due to these effects in any single structure;
and for the \citetalias{Alcock1979} test
\citep{Sutter2014AP, Hamaus2020}.

\subsubsection{Stacked radial density profiles}

We stack all of the voids in the simulation to produce averaged radial profiles for the density contrast in the voids.
The top panels of Fig.~\ref{fig:radial_density} 
show these profiles, normalised by each void's
effective radius,
for the \perhMpc{4} and \perhMpc{12} resolution simulations, respectively. 
The shaded areas of these figures
show the density contrast of the middle
~68.5\% 
and ~95\% 
of cells at each rescaled radial distance.
These shaded bands show that walls begin appearing in some directions even when $r<\reff$, where the average density contrast is still negative.
Likewise, the overcompensation continues much further away from the centre than $\reff$, continuing to be visible past $r=2\reff$.

This is a natural consequence of the fact that
the voids are not all close to spherical.
The lower panels in these figures show this explicitly,
showing what proportion of points at a given distance are inside of the void or outside of it, either in the walls or in a different void (solid curve).
For comparison, these panels also show the case of spheres at infinite resolution (dotted line; a step function)
and a set of spheres approximated on the grid with
radius 3--5 grid cells (dashed curve) -- comparable to the effective radius of typical voids in these simulations.
The latter is not a step function, due to the spheres being represented on the finite-resolution grid.
However, the transition between cells inside and outside of the void occurs over a shorter range of radial distances for spheres on the grid as compared to the simulated voids,
confirming that the simulated voids are generally not spherical.

The mean density contrast at void centres is 
$\delta\approx-0.69$ at \perhMpc{4} resolution 
and $\delta\approx-0.42$ at \perhMpc{12} resolution.
As expected, the magnitude of the density contrast --
minimum, maximum, and average --
decreases as the
the smoothing scale becomes larger. 
In both simulations the density contrast decays to the whole-volume mean of $\delta=0$ as $r$ becomes several times larger than $\reff$,
but the rate at which this occurs differs with smoothing scale.
For \perhMpc{12} resolution, the mean $\delta$ over the shell at distance $r=2\reff$ is approximately $0.01$,
while at \perhMpc{4} resolution, the distance necessary to reach the same $\delta\approx0.01$ is roughly $r=3\reff$ 
(which is physically smaller, as the voids at that resolution are smaller).

\subsubsection{Fitting HSW profiles}

\begin{figure}
	\includegraphics{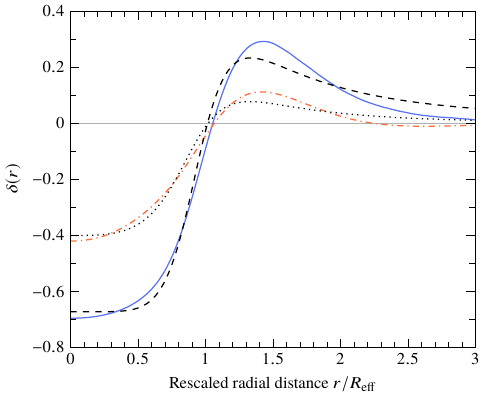}
    \caption{
    The stacked radial density contrast profiles from Fig.~\ref{fig:radial_density} and the corresponding fitted HSW profiles. Namely, the \perhMpc{4} simulation density contrast (solid blue), its fitted HSW profile (dashed black),
    the \perhMpc{12} simulation density contrast (dash-dotted orange), and its fitted HSW profile (dotted black).
    These HSW profiles use \eqref{eq:hsw}, fitting only the typical four parameters.
    }
    \label{fig:hsw_poor_fit}
\end{figure}
\begin{figure}
	\includegraphics{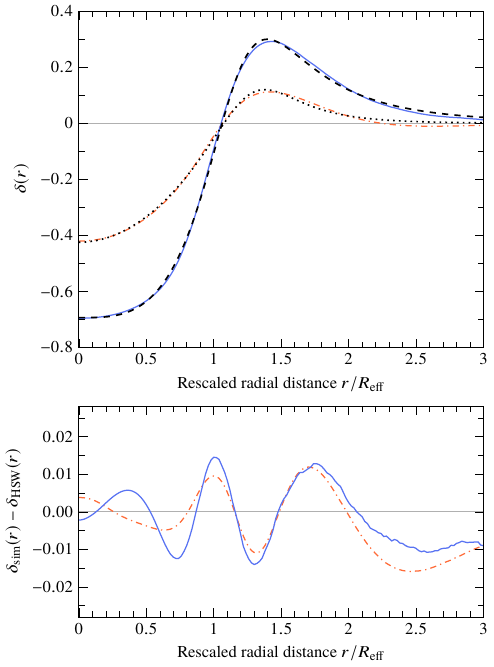}
    \caption{
    Upper: The stacked radial density contrast profiles from Fig.~\ref{fig:radial_density} and the corresponding fitted HSW profiles. Namely, the \perhMpc{4} simulation density contrast (solid blue), its fitted HSW profile (dashed black),
    the \perhMpc{12} simulation density contrast (dash-dotted orange), and its fitted HSW profile (dotted black).
    These HSW profiles use \eqref{eq:hsw2},
    with the fifth fitted parameter $\HSWradius$ in place of $\reff$ to determine the profile. The $x$-axis is normalised by the original $\reff$.
    Lower: Difference between HSW density contrast and stacked radial density contrast for the \perhMpc{4} resolution simulation (solid) and the \perhMpc{12} resolution simulation (dash-dotted).
    }
    \label{fig:hsw}
\end{figure}

Fig.~\ref{fig:hsw_poor_fit} shows the HSW universal radial density profile in \eqref{eq:hsw} fitted to the stacked radial profiles in the two simulations.
The solid blue curve shows the observed density contrast in the \perhMpc{4} resolution simulation,
and the dashed black curve shows the HSW profile fitted to it.
The dot-dashed orange curve and dotted black curve, respectively, show the same for the \perhMpc{12} resolution simulation.
The fitted parameters are given in Table~\ref{tab:hsw_params}, in the ``using $\reff$'' rows.
These parameters give poor fits for the density contrast near the peak in the wall and further outside the void, and fit a flatter bottom to the void than profiles of the simulated voids.

The HSW profile has a dependence on the effective radius $\reff$.
The same $\reff$ computed from \eqref{eq:reff} and used to 
normalise each void radius when stacking voids
is used in \eqref{eq:hsw} 
to fit the HSW profile.
As this produces a poor fit for our data,
we also try fitting it as a free parameter.
We adjust \eqref{eq:hsw} to
\begin{equation}
\label{eq:hsw2}
    \delta = \delta_c\frac{1-(r/\scaleradius)^\alpha}{1+(r/\HSWradius)^\beta},
\end{equation}
where $\HSWradius$ is a parameter that determines the size of the void in place of $\reff$.
The rows of Table~\ref{tab:hsw_params} with ``fitting $\HSWradius$'' list the adjusted parameter sets.
The fitted $\HSWradius$ is 25\% and 34\% larger than the effective radius for \perhMpc{4} and \perhMpc{12} resolution, respectively.
Fig.~\ref{fig:hsw} shows the HSW profiles with these new parameters (curve colours and styles matching Fig.~\ref{fig:hsw_poor_fit}), 
and the error $\delta_\mathrm{sim} - \delta_\mathrm{HSW}$ (lower panel), which is now visibly much smaller. 

\begin{table}
    \centering
    \caption{Values for the fitted parameters for the HSW profiles for both simulations.}
    \label{tab:hsw_params}
    \begin{tabular}{lrrrrr} 
        \hline
         Resolution & $\delta_c$ & $\scaleradius$ & $\HSWradius$ & $\alpha$ & $\beta$\\
        \hline
        \perhMpc{4}, using $\reff$ & -0.673 & 1.02 & $\reff$ & 5.79 & 8.02\\
        \perhMpc{4}, fitting $\HSWradius$ & -0.693 & 1.06 & $1.25\,\reff$ & 3.31 & 7.85\\
        \perhMpc{12}, using $\reff$ & -0.400 & 1.02 & $\reff$ & 3.39 & 6.60\\
        \perhMpc{12}, fitting $\HSWradius$ & -0.424 & 1.08 & $1.34\,\reff$ & 2.01 & 9.17\\
        \hline
    \end{tabular}
\end{table}

We interpret this as showing that our measure of effective radius is not interchangeable with that of others who fit HSW profiles.
That is, our value $\HSWradius$ is more comparable to the meaning of ``effective radius'' as used in a $N$-body simulation or observations
\citep[e.g.,][]{Hamaus2014}
than our $\reff$ is.
This difference in notions of effective radius may be a resolution effect, due to our use of the continuous fluid approximation.
The large physical size of the grid cells as compared to structures means that thinnest possible wall in the simulations
is much thicker than a wall observed in data or one in an $N$-body simulation.
This is supported by the fact that our best-fit $\HSWradius$ values are converging towards $\reff$ as we increase physical resolution (Table~\ref{tab:hsw_params}).
Note that a further increase in resolution is unachievable in practice due to the limits of the fluid approximation.

This exclusion of over-dense cells from voids may be another cause for this difference,
as it causes the peak in density to occur further from $r=\reff$.
For a direct comparison to results from Newtonian $N$-body simulations -- for which the HSW profiles provide a very good fit -- we would need to simulate collisionless particles alongside NR.

In light of this,
it may be more meaningful to rescale the void size functions of Section \ref{sec:vsf}
such that the ``effective radius'' giving the ``size'' of the void is $\HSWradius=1.25\reff$ or $1.34\reff$ 
(Table~\ref{tab:hsw_params})
rather than the effective radius $\reff$ itself as we define it in \eqref{eq:reff}, 
or in some other way compensate for the decrease in volume that results from low physical resolution.

\subsubsection{General-relativistic radial profiles}
\label{sec:curvatureProfiles}

\begin{figure}
	\includegraphics{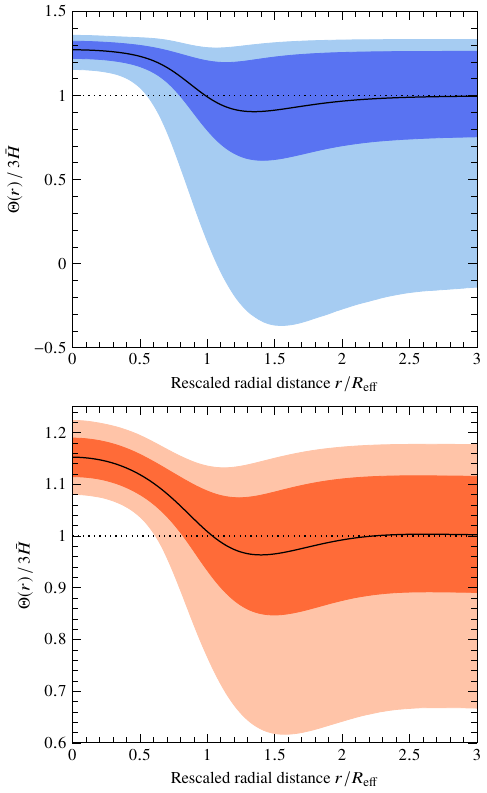}
    \caption{
    The average fluid-intrinsic expansion $\Theta$ on a spherical shell at distance $r$ from the stacked void centres,
    with each void scaled by its effective radius $\reff$.
    The shaded areas show the middle ~68.5\% and ~95\% of values for expansion on that shell.
    A horizontal line shows the whole-box average expansion $\Theta = 3\Hall$.
    The upper panel shows the \perhMpc{4} resolution simulation, the lower, \perhMpc{12}.
    }
    \label{fig:radial_theta}
\end{figure}
\begin{figure}
	\includegraphics{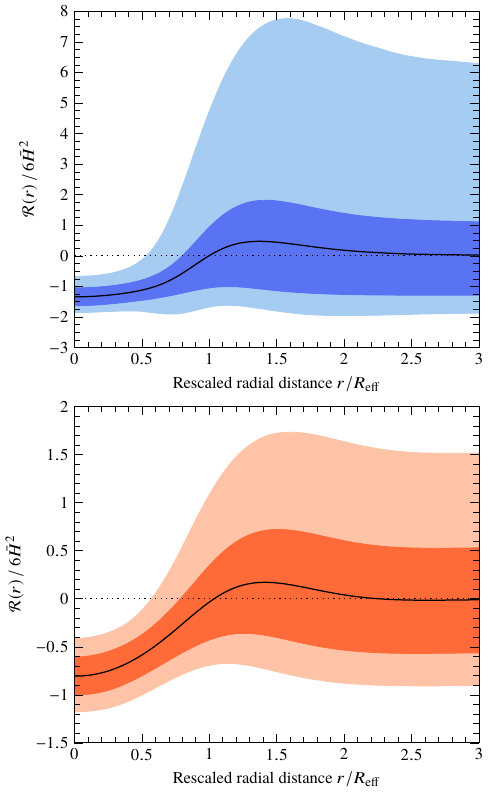}
    \caption{
    Upper:
    The average fluid-intrinsic curvature scalar $\threericci$ on a spherical shell at distance $r$ from the stacked void centres,
    with each void scaled by its effective radius $\reff$. 
    The shaded areas show the middle ~68.5\% and ~95\% of values for the Ricci scalar on that shell.
    The upper panel shows the \perhMpc{4} resolution simulation, the lower, \perhMpc{12}.
    }
    \label{fig:radial_ricci}
\end{figure}
\begin{figure}
	\includegraphics{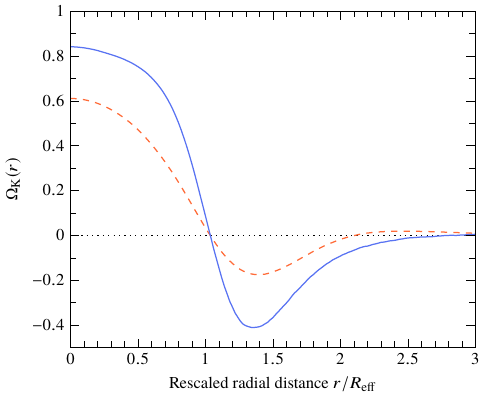}
    \caption{
    Median $\OK$ over all voids for each distance $r/\reff$. For each void, the domain of $\OK$ is a shell of radius $r$ centred on the void's centre, with the void scaled by its effective radius $\reff$.
    The solid curve shows the median for voids in the \perhMpc{4} resolution simulation, the dashed curve, \perhMpc{12}.}
    \label{fig:radial_omega}
\end{figure}

We also produce
stacked radial profiles for three general-relativistic scalars,
which are important in the Buchert averaging scheme \BMR{}.
To obtain dimensionless profiles (analogous to obtaining a profile for $\delta$)
we normalise the expansion and curvature scalar by the effective Hubble rate.
The effective Hubble function $\HD$ 
(over a spatial domain $\mathcal{D}$)
is
defined in terms of the average expansion as
\begin{equation}
\label{eq:Hubble}
    \HD:=\frac{\dot{a}_\mathcal{D}}{a_\mathcal{D}}=\frac{1}{3}\langle\Theta\rangle_\mathcal{D}\,.
\end{equation}
Thus the effective scale factor $a_\mathcal{D}$ is defined in terms of the evolution of the fluid volume $\mathcal{V}_\mathcal{D}(t)$ compared to its initial volume $\mathcal{V}_{\mathcal{D}_i}$
\begin{equation}
\label{eq:scalefactor}
    a_\mathcal{D}(t) = \left(\frac{\mathcal{V}_\mathcal{D}(t)}{\mathcal{V}_{\mathcal{D}_i}}\right)^\frac{1}{3},
\end{equation}
$\dot{a}_\mathcal{D}=\mathrm{d}a_\mathcal{D}/\mathrm{d}t$, and the volume is defined in \eqref{eq:cell_volume}. 
The average scale factor over the entire simulation volume is denoted $\bar a$, and $\Hall= \bar{\Theta}/3$.

The first scalar we
study is 
the expansion scalar 
in the rest frame of the fluid,
$\Theta\equiv \nabla_\mu u^\mu$, where $\nabla_\mu$ is the covariant derivative associated with $g_{\mu\nu}$ and $u^\mu$ is the 4-velocity of the fluid. We calculate $\Theta$ (and $\Hall$ and the Ricci scalar) using existing routines in \texttt{mescaline} \citep{Macpherson2019}.

Fig.~\ref{fig:radial_theta} shows the stacked spherical profiles  of $\Theta$ (solid curve)
in the simulations at \perhMpc{4} (upper panel) and \perhMpc{12} (lower panel) resolution.
In both panels, the shaded regions show the middle ~68.5\% and ~95\%
of $\Theta$ values on each spherical shell.
In the \perhMpc{4} resolution simulation,
we find that in the centres of the voids
the mean value for $\Theta$ is $1.27$ times the large-scale average of $3\Hall$,
or $1.15 \times 3\Hall$ at \perhMpc{12} resolution.

This explicitly shows that voids expand at least 
$\sim$10--30\% faster than the large-scale average. 
This is in agreement with predictions from linear perturbation theory \citep[e.g.,][]{Lahav1991} and previous NR studies \citep[e.g., Fig.~10 of][]{Macpherson2019}. 
The average expansion rate is lower on shells that intersect the walls,
to a minimum shell average of 91\% and 97\% (\perhMpc{4} resolution and \perhMpc{12}, respectively) of the large-scale average.
In some cases individual cells have negative $\Theta$, implying collapse. 
Our expansion profiles here can be considered as lower limits on the expansion in the centre and walls of voids, due to the coarse smoothing scales in our simulations.

We also study the fluid-intrinsic curvature scalar $\threericci$ (defined in equation (4.15) of \BMRt{}).
This can be roughly considered the 3-Ricci scalar in the rest frame of the fluid. 
Normalising this by $6\Hall^2$ gives a dimensionless quantity.
Fig.~\ref{fig:radial_ricci} shows the stacked spherical profiles  of $\threericci$ (solid curve)
in the simulations at \perhMpc{4} (upper panel) and \perhMpc{12} (lower panel) resolution.
In both panels, the shaded regions show the middle ~68.5\% and ~95\%
of $\threericci$ values on each spherical shell.
We see that voids in our simulations have negative spatial curvature, while in the walls the spatial curvature is positive.
As the distance from the void increases beyond the wall, the average 
spatial curvature tends towards 
the whole-volume average spatial curvature \citep[flat; see][]{Macpherson2019}.

The third scalar we study here is $\OmegaCurvature$ \citep{Buchert2000}, the kinetic curvature parameter
\begin{equation}
\label{eq:omega_curvature}
    \OmegaCurvature = -\frac{\langle\threericci\rangle_\mathcal{D}}{6 \HD^2} = -\frac{3\langle\threericci\rangle_\mathcal{D}}{2\langle\Theta\rangle_\mathcal{D}^2}\,.
\end{equation}
Buchert introduces a general formalism for spatial averages on arbitrary domains, $\mathcal{D}$, which leads to markedly different results depending on the domain of averaging. To study the effect of small scale inhomogeneities, the domain $\mathcal D$ is that of our fluid elements as limited by the resolution scale of the simulation. The procedure of stacking voids over the entire box is then an operational realisation of how the volume average void fraction is defined in the timescape interpretation of the Buchert scheme \citep{wiltshireCosmicClocks2007a,wiltshireAverageObservationalQuantities2009}.

For our radial profiles, the averaging domain is 
an individual spherical shell.
Thus, we find a $\OK$ profile for a single void by finding the average values $\Theta$ and $\threericci$ on individual shells of radius $r$ around that void. 
The profile for each void is then rescaled by the void's effective radius $\reff$ and the median $\OK$ of the set of all void profiles is found at each distance $r/\reff$ to produce the stacked profile, shown in Fig.~\ref{fig:radial_omega}.

The magnitude of $\OK$ reaches 0.8 in the void centre for the smallest physical resolution we study here. We expect this to be even larger for a situation with resolution below \perhMpc{4} (i.e., an $N$-body simulation or galaxy data).
While the negative curvature (Fig.~\ref{fig:radial_omega} and Fig.~\ref{fig:radial_ricci}) and higher expansion (Fig.~\ref{fig:radial_theta}) in void centres are expected in 
GR, this work is the first time it has been shown explicitly using realistic simulations in NR \citep[though see][]{Macpherson2018}.

\subsection{Boundary distance profiles}

The over-dense walls of voids tend to be thin compared to the size of the average void.
For example, at \perhMpc{4} resolution,
we find that most walls are only a single \perhMpc{4} cell thick.
However, the stacked radial void profiles for this simulation in the previous section have a broad over-density.
The reason that these stacked radial profiles have a much lower mean density contrast than that of the typical wall,
along with a much greater width,
is because
voids are not spherical \citep{Cautun2014, Cautun2016},
as in the cosmic web their shape can be constrained by the surrounding voids.

On a spherical shell with a radius near the effective radius,
some positions will still be within the void, and be highly under-dense,
while others will be within the wall, and highly over-dense,
and others will be outside of the void entirely,
and have a density which averages to be close to the overall average density.
The lower plot of Fig.~\ref{fig:radial_density}
shows the mixture of void cells and non-void cells averaged over at each radial distance.
Due to this, the added density of the walls is spread out over a large radial distance.
\citet{Schuster2023} confirm that the most spherical voids have sharper walls when stacked, as a consequence of this.

\citet{Cautun2014,Cautun2016} propose another method of constructing averaged void profiles which does not assume they are spherically symmetric.
Rather than measuring distance from the centre of the void,
(additionally complicated by having to choose a centre, see Section~\ref{sec:centres}),
they suggest constructing a profile as a function of distance from the void's boundary.
This ensures that at a given distance 
the average is taken only over points inside, only outside, or only on the boundary of the void, 
avoiding mixing these cases.

We also produce profiles using this approach, using the set of centres of wall cells adjacent to the interior cells of a given void as the boundary for that void, and define distance from the boundary as the distance to the closest point in that set. 
We then produce an average of the profiles of 2000 randomly-selected voids (rather than all of them, for computational reasons).

\begin{figure}
	\includegraphics{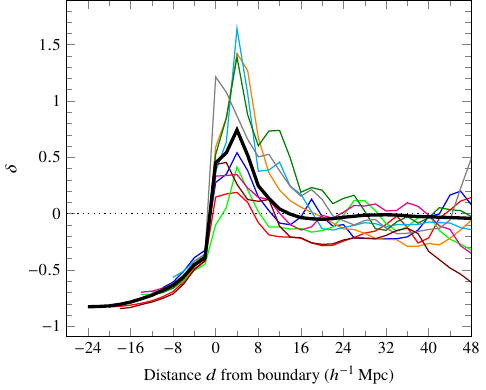}
    \caption{
    The average density contrast at a distance $d$ from the boundary in each of 10 randomly chosen voids (10 coloured curves) with the 2000-void average (thicker black), all in the \perhMpc{4} resolution simulation. Negative distances are in the interior of the void.
    }
    \label{fig:boundary_example}
\end{figure}

Fig.~\ref{fig:boundary_example} shows some examples of individual void boundary profiles for density,
along with the 2000-void average.
The ten coloured curves show the density of ten randomly selected voids in the \perhMpc{4} resolution simulation
as a function of distance from the void's boundary,
and the thicker black curve is the average over 2000 such profiles.
For these profiles, there is no need to rescale by the effective radius to align the walls.
Note that this means that profiles of smaller voids do not go as far to the left on the $x$-axis.

\begin{figure}
	\includegraphics{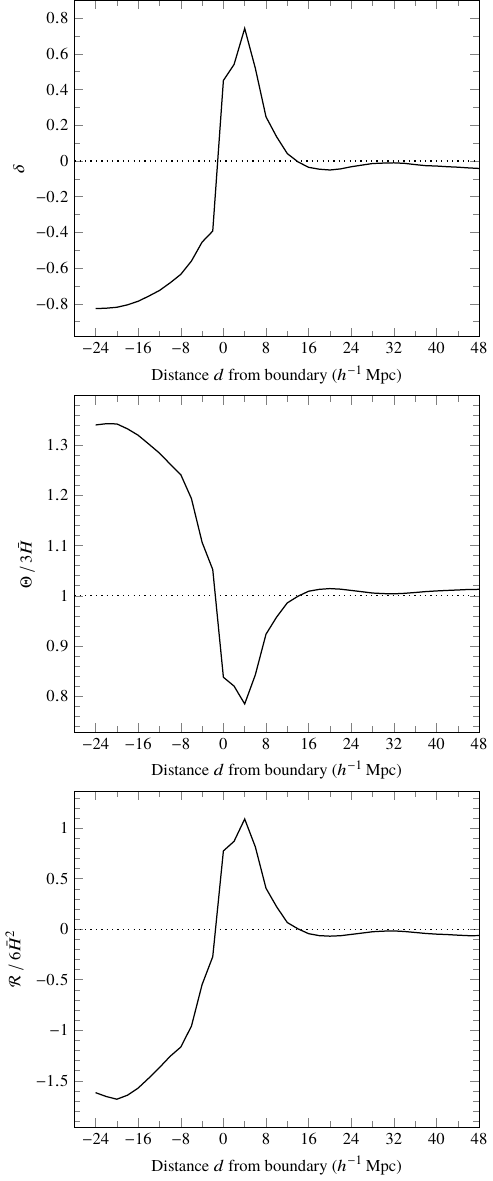}
    \caption{
    The average values of $\delta$, $\Theta$, and $\threericci$ (top to bottom) on a shell at distance $d$ from the boundary of the void, averaged over 2000 voids from the \perhMpc{4} resolution simulation.
    Negative distances are in the interior of the void.
    }
    \label{fig:boundary_4}
\end{figure}

\begin{figure}
	\includegraphics{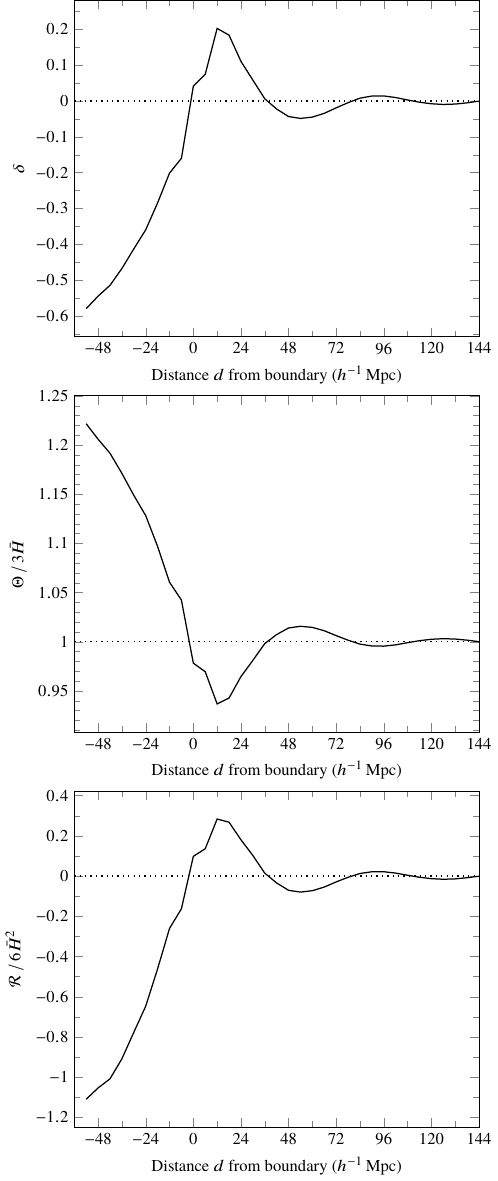}
    \caption{
    The average values of $\delta$, $\Theta$, and $\threericci$ (top to bottom) on a shell at distance $d$ from the boundary of the void, averaged over 2000 voids from the \perhMpc{12} resolution simulation.
    Negative distances are in the interior of the void.
    }
    \label{fig:boundary_12}
\end{figure}

Figs.~\ref{fig:boundary_4} and \ref{fig:boundary_12} show the averaged profile of $\delta$ top panels), $\Theta$ (middle panels), and $\threericci$ (bottom panels) in the two simulations as a function of distance from the void boundary.
The extremum that appears in the walls for each variable has greater magnitude than for the corresponding stacked spherical profile.
Likewise, the gradient near the walls is greater.
These properties are to be expected, as the walls are all aligned at distance $d=0$, rather than scattered at varying radial distances around $r=\reff$.

Another phenomenon seen in these profiles which cannot be seen in the stacked spherical profiles is that outside the void, beyond the walls,
there is a second range of under-density followed by a second peak.
This effect is more noticeable in the \perhMpc{12} resolution simulation.
This is due to the increased likelihood of finding another void after passing through a wall, as the walls are thin.
The distance between peaks is then related to typical void size,
as it requires moving through another void (not necessarily along its ``diameter'') to reach another wall.
In the stacked spherical profiles,
the void walls are spread over a wide range of radial distances,
and so obscure this phenomenon.

\section{Conclusions}
\label{sec:conclusions}
In this work, we explored void statistics in non-linear general relativity for the first time.
We have implemented the \textsc{Watershed Einstein Topography} (WET) void finder
for use with simulation data output by the \einsteintoolkit{}.
Our simulations approximate matter as a
continuous fluid,
which necessitates differences in the implementation of this void finder
compared to publicly-available void finders.

Having identified the voids in two simulations of different physical resolution, 
we studied
the void size function,
void fraction,
and stacked radial density profiles.
We fit HSW profiles \citep{Hamaus2014} to
our stacked radial density 
profiles for each simulation.
As discussed in Section~\ref{sec:radialProfiles},
we alter this process by fitting a parameter $\HSWradius$ in place of the effective radius $\reff$.
This is because our calculated $\reff$ is not necessarily equivalent to the measure of effective radius typically used for HSW profiles in other works, likely due to our comparably low physical resolution or exclusion of over-dense cells.

We also utilise the boundary distance profile approach of \citet{Cautun2016} 
to produce a second set of profiles for each quantity.
Using this approach, the distinction between cells in the interior of the void, within the walls, and beyond the walls is clearer.
We find that when at a distance from the walls equal to roughly the typical effective radius of a void,
the density once again drops lower than the large-scale average, then has a second over-dense peak.
This is
expected
because the walls are thin,
so typically passing through a wall will lead to another void.
It is not possible to see this feature in the stacked spherical profiles,
as the void walls are spread over a wide range of radial distances,
since the voids are not spherical.

In addition to the statistics usually obtained from $N$-body simulations or observation,
we studied 
a set of general-relativistic void statistics.
In particular, the fluid-intrinsic expansion $\Theta$,
the fluid-intrinsic curvature scalar $\threericci$,
and the kinetic curvature parameter $\OK$. 
We found that, as expected, these quantities vary within the void.
Numerical relativity simulations such as 
those we use here allow us to study the general-relativistic qualities of voids without imposing symmetries 
on the metric tensor.
The expansion rate is, on average, 27\% greater than the large-scale average of $\bar{\Theta} = 3\Hall$ at the centre of voids in the \perhMpc{4} resolution simulation (15\% in the \perhMpc{12} resolution simulation),
and reaches a minimum average value of 97\% (91\%) of the large-scale average in the \perhMpc{4} (\perhMpc{12}) resolution simulation.
We find that voids are negatively curved, with the kinetic curvature parameter $\OK$ being 0.83 and 0.61 at the centre of voids at \perhMpc{4} and \perhMpc{12} resolution, respectively,
while the walls are positively curved, with $\OK$ of -0.58 and -0.18, respectively. 
We would reasonably expect these values to get even larger in magnitude with a higher (and thus more realistic) physical resolution.
Our results thus represent a lower limit on the expansion and curvature of voids in a matter-dominated model universe.
This is due to the limitation of our simulations using a continuous fluid approximation; which restricts the smallest physical scales we can sample.

While NR simulations in cosmology are not new, 
being able to quantify small-scale expansion and curvature in physically relevant regions -- i.e., cosmic voids -- required the development of appropriate void-finding tools.
Although $\OK$ is large near void centres,
if one takes averages over randomly chosen regions that contain both voids and walls,
its magnitude is small \citep{Macpherson2019}.
This can be seen at large values of $r/\reff$ in the radial profiles and large distances from the void boundary in the boundary distance profiles.
It is necessary to unambiguously identify void regions, as we have done here, to see this signal clearly 
in averages on these scales.

The  consistency of Fig.~\ref{fig:radial_omega} with the kinetic curvature interpretation of cosmic voids raises important questions about distinguishing peculiar velocities from the background and is central to many debates in observational cosmology.
These questions raise the possibility that large angle anomalies observed in the CMB and whole sky catalogues of distant sources \citep[for a recent review, see][]{Aluri2023} might be resolved within general relativity. 
Work is in progress to address this by examining the detailed properties of
the structure dipole in our simulations (Williams et al., in prep).

We emphasise that most cosmological simulations enforce an FLRW expansion of space-time throughout the simulation domain, serving as checks of the standard cosmology.
Analytic perturbation methods on FLRW backgrounds have 
identified gauges in which Newtonian $N$-body results can be embedded \citep{Fidler_2016,Fidler_2017}, as well as consistency with post-Newtonian results 
in the presence of large density contrasts \citep{Clifton_2020}. However, FLRW has at least three distinct notions of homogeneity and isotropy corresponding to the classic foliation and gauge choices of \citet{Bardeen_1980} and their generalizations \citep{Bicak_2007}. Possible differences from FLRW expectations are therefore not surprising once strict homogeneity and isotropy of the average evolution is relaxed.

In this work we have explicitly demonstrated the spatial fluctuations of expansion and curvature in void regions in GR. 
However, the impact of these variations remains to be thoroughly explored. While we find qualitatively similar density profiles to Newtonian results, we have not performed a direct comparison to an equivalent Newtonian model universe.
A strict comparison would require a treatment of matter as collisionless particles alongside NR (or a comparison with a similar fluid-based Newtonian code) to minimise numerical artefacts in the comparison. 
The ET has recently been combined with 
SPH \citep{Magnall2023}, however, this adaptation is still being tested for simulations with realistic initial data. 

While some relativistic effects have been calculated from purely Newtonian simulations \citep[e.g.,][]{Milillo2015,Thomas2015b,Borzyszkowski2017} -- including some which vanish in the Newtonian limit \citep[e.g.,][]{Bruni2014,Thomas2015,Tram2019} -- curvature and expansion profiles inside voids have not yet been studied. Further, such variations in curvature are unable to impact the large-scale expansion of Newtonian simulations due to the implementation of the Friedmann equation. 
However, calculating curvature and expansion profiles in post-processing is in principle straightforward and it would be interesting to explore how well such profiles can be reliably predicted from density profiles alone. Such an analysis could allow for the generation of general-relativistic profiles from Newtonian $N$-body simulations or observational catalogues. However, this would first require a rigorous comparison of our density profiles to a Newtonian analogue.

The magnitude of $\OK$ we find in void centres is of the same order as that found by \citet[table~1, fig.~1]{DuleyTimescapeRadiation2013} for the timescape model fit to the angular scales of the CMB acoustic peaks in the 2013 Planck data release. Since more than one time parameter is associated with the present epoch in the timescape model, these being related to particular averaging domains, comparing NR simulations with observation require subtle reinterpretation of standard procedures. Consistency with previous analyses is likely to involve a reinterpretation of the time-dependent reparameterization of \citet[Sec. 5.3]{Adamek2016} and is left to future work (Wiltshire et al., in prep). Such studies also necessitate careful analysis of the backreaction density, $\OQ$. Previous analysis found that it cancelled well below the numerical error for averages over the whole simulation box \citep{Macpherson2019}. This may be related to the periodic boundary conditions enforcing a torus topology of the spatial hypersurfaces. The impact of this condition on the size of the backreaction effect has not been studied in GR. However, it is well-known to force backreaction 
to zero on large scales in Newtonian gravity \citep{Buchert1997}.

In addition to our approximation of a continuous fluid and periodic boundary conditions, the most important caveats to our results are as follows.
We use data from a single spatial slice of each simulation to find the voids.
While this is common for void-finding in simulated data, it limits our ability to compare with observation,
as voids are observed on our past light cone. 
General-relativistic ray tracing in 
simulations like those presented here \citep{Macpherson2023}
would enable us to find the voids in redshift space, 
allowing for a more direct comparison to observation.
Secondly, our simulations omit a cosmological constant in the evolution. 
This influences the growth of structures in the model universe, typically resulting in larger density contrasts than a comparable model with $\Lambda$. Such an omission will impact the void statistics in the simulation.
We are working on implementing this into the \einsteintoolkit{}, 
which will allow for closer comparison to Newtonian $N$-body simulations.

\section*{Acknowledgements}

We thank the anonymous referee, whose careful reading improved the clarity of our manuscript.
We also thank Nico Hamaus for helpful discussion and comments, and members of the UC Gravity and Cosmology group for their feedback and suggestions. 
Support for HJM was provided by NASA through the NASA Hubble Fellowship grant HST-HF2-51514.001-A awarded by the Space Telescope Science Institute, which is operated by the Association of Universities for Research in Astronomy, Inc., for NASA, under contract NAS5-26555. DLW and CS are supported by the Marsden Fund administered by the Royal Society of New Zealand, Te Apārangi under grants M1271 and M1278. 
The simulations used in this work were performed on the DiRAC@Durham facility managed by the Institute for Computational Cosmology on behalf of the STFC DiRAC HPC Facility (www.dirac.ac.uk). The equipment was funded by BEIS capital funding via STFC capital grants ST/P002293/1, ST/R002371/1 and ST/S002502/1, Durham University and STFC operations grant ST/R000832/1. DiRAC is part of the National e-Infrastructure.
We acknowledge the use of the University of Canterbury's Gravity cluster,
an inter-faculty HPC resource shared between the School of Mathematics and Statistics and the School of Physical and Chemical Sciences, for the post-processing of simulation data.

\section*{Data Availability}

The void catalogue for the two simulations used in this work is available at
\url{https://github.com/mwilliamsnz/NR-void-catalogue}. 
The simulation data, void-finding code, and other post-processing scripts used to produce this data are available upon reasonable request to the corresponding author.



\bibliographystyle{mnras}
\bibliography{bibliography}

\begin{thebibliography}{}
\makeatletter
\relax
\def\mn@urlcharsother{\let\do\@makeother \do\$\do\&\do\#\do\^\do\_\do\%\do\~}
\def\mn@doi{\begingroup\mn@urlcharsother \@ifnextchar [ {\mn@doi@}
  {\mn@doi@[]}}
\def\mn@doi@[#1]#2{\def\@tempa{#1}\ifx\@tempa\@empty \href
  {http://dx.doi.org/#2} {doi:#2}\else \href {http://dx.doi.org/#2} {#1}\fi
  \endgroup}
\def\mn@eprint#1#2{\mn@eprint@#1:#2::\@nil}
\def\mn@eprint@arXiv#1{\href {http://arxiv.org/abs/#1} {{\tt arXiv:#1}}}
\def\mn@eprint@dblp#1{\href {http://dblp.uni-trier.de/rec/bibtex/#1.xml}
  {dblp:#1}}
\def\mn@eprint@#1:#2:#3:#4\@nil{\def\@tempa {#1}\def\@tempb {#2}\def\@tempc
  {#3}\ifx \@tempc \@empty \let \@tempc \@tempb \let \@tempb \@tempa \fi \ifx
  \@tempb \@empty \def\@tempb {arXiv}\fi \@ifundefined
  {mn@eprint@\@tempb}{\@tempb:\@tempc}{\expandafter \expandafter \csname
  mn@eprint@\@tempb\endcsname \expandafter{\@tempc}}}

\bibitem[\protect\citeauthoryear{Adamek, Daverio, Durrer  \& Kunz}{Adamek
  et~al.}{2016}]{Adamek2016}
Adamek J.,  Daverio D.,  Durrer R.,   Kunz M.,  2016, \mn@doi [\jcap]
  {10.1088/1475-7516/2016/07/053}, {\protect\rm07}, 053

\bibitem[\protect\citeauthoryear{{Adamek}, {Barrera-Hinojosa}, {Bruni}, {Li},
  {Macpherson}  \& {Mertens}}{{Adamek} et~al.}{2020}]{Adamek2020}
{Adamek} J.,  {Barrera-Hinojosa} C.,  {Bruni} M.,  {Li} B.,  {Macpherson}
  H.~J.,   {Mertens} J.~B.,  2020, \mn@doi [Classical and Quantum Gravity]
  {10.1088/1361-6382/ab939b}, \href
  {https://ui.adsabs.harvard.edu/abs/2020CQGra..37o4001A} {37, 154001}

\bibitem[\protect\citeauthoryear{{Adamek}, {Clarkson}, {Durrer}, {Heinesen},
  {Kunz}  \& {Macpherson}}{{Adamek} et~al.}{2024}]{Adamek2024}
{Adamek} J.,  {Clarkson} C.,  {Durrer} R.,  {Heinesen} A.,  {Kunz} M.,
  {Macpherson} H.~J.,  2024, \mn@doi [arXiv e-prints]
  {10.48550/arXiv.2402.12165}, \href
  {https://ui.adsabs.harvard.edu/abs/2024arXiv240212165A} {p. arXiv:2402.12165}

\bibitem[\protect\citeauthoryear{Alcock \& Paczy\'nski}{Alcock \&
  Paczy\'nski}{1979}]{Alcock1979}
Alcock C.,  Paczy\'nski B.,  1979, \mn@doi [Nature]
  {https://doi.org/10.1038/281358a0}, 281, 358

\bibitem[\protect\citeauthoryear{{Aluri} et~al.,}{{Aluri}
  et~al.}{2023}]{Aluri2023}
{Aluri} P.~K.,  et~al., 2023, \mn@doi [Classical and Quantum Gravity]
  {10.1088/1361-6382/acbefc}, \href
  {https://ui.adsabs.harvard.edu/abs/2023CQGra..40i4001K} {40, 094001}

\bibitem[\protect\citeauthoryear{{Baiotti}, {Hawke}, {Montero}, {L{\"o}ffler},
  {Rezzolla}, {Stergioulas}, {Font}  \& {Seidel}}{{Baiotti}
  et~al.}{2005}]{Baiotti:2005GRHydro}
{Baiotti} L.,  {Hawke} I.,  {Montero} P.~J.,  {L{\"o}ffler} F.,  {Rezzolla} L.,
   {Stergioulas} N.,  {Font} J.~A.,   {Seidel} E.,  2005, \mn@doi [\prd]
  {10.1103/PhysRevD.71.024035}, \href
  {https://ui.adsabs.harvard.edu/abs/2005PhRvD..71b4035B} {71, 024035}

\bibitem[\protect\citeauthoryear{{Baiotti}, {Giacomazzo}  \&
  {Rezzolla}}{{Baiotti} et~al.}{2008}]{Baiotti:2008}
{Baiotti} L.,  {Giacomazzo} B.,   {Rezzolla} L.,  2008, \mn@doi [\prd]
  {10.1103/PhysRevD.78.084033}, \href
  {https://ui.adsabs.harvard.edu/abs/2008PhRvD..78h4033B} {78, 084033}

\bibitem[\protect\citeauthoryear{{Baker}, {Centrella}, {Choi}, {Koppitz}  \&
  {van Meter}}{{Baker} et~al.}{2006}]{Baker:2006}
{Baker} J.~G.,  {Centrella} J.,  {Choi} D.-I.,  {Koppitz} M.,   {van Meter} J.,
   2006, \mn@doi [\prl] {10.1103/PhysRevLett.96.111102}, \href
  {https://ui.adsabs.harvard.edu/abs/2006PhRvL..96k1102B} {96, 111102}

\bibitem[\protect\citeauthoryear{{Bardeen}}{{Bardeen}}{1980}]{Bardeen_1980}
{Bardeen} J.~M.,  1980, \mn@doi [\prd] {10.1103/PhysRevD.22.1882}, \href
  {https://ui.adsabs.harvard.edu/abs/1980PhRvD..22.1882B} {22, 1882}

\bibitem[\protect\citeauthoryear{{Barrera-Hinojosa} \& {Li}}{{Barrera-Hinojosa}
  \& {Li}}{2020a}]{gramses1}
{Barrera-Hinojosa} C.,  {Li} B.,  2020a, \mn@doi [\jcap]
  {10.1088/1475-7516/2020/01/007}, \href
  {https://ui.adsabs.harvard.edu/abs/2020JCAP...01..007B} {{\protect\rm01},
  007}

\bibitem[\protect\citeauthoryear{{Barrera-Hinojosa} \& {Li}}{{Barrera-Hinojosa}
  \& {Li}}{2020b}]{gramses2}
{Barrera-Hinojosa} C.,  {Li} B.,  2020b, \mn@doi [\jcap]
  {10.1088/1475-7516/2020/04/056}, \href
  {https://ui.adsabs.harvard.edu/abs/2020JCAP...04..056B} {{\protect\rm04},
  056}

\bibitem[\protect\citeauthoryear{Baumgarte \& Shapiro}{Baumgarte \&
  Shapiro}{1998}]{Baumgarte1998}
Baumgarte T.~W.,  Shapiro S.~L.,  1998, \mn@doi [\prd]
  {10.1103/PhysRevD.59.024007}, 59, 024007

\bibitem[\protect\citeauthoryear{{Beck}, {Csabai}, {R{\'a}cz}  \&
  {Szapudi}}{{Beck} et~al.}{2018}]{Beck_2018}
{Beck} R.,  {Csabai} I.,  {R{\'a}cz} G.,   {Szapudi} I.,  2018, \mn@doi
  [\mnras] {10.1093/mnras/sty1688}, \href
  {https://ui.adsabs.harvard.edu/abs/2018MNRAS.479.3582B} {479, 3582}

\bibitem[\protect\citeauthoryear{Bentivegna \& Bruni}{Bentivegna \&
  Bruni}{2016}]{Bentivegna2016}
Bentivegna E.,  Bruni M.,  2016, \mn@doi [\prl]
  {10.1103/PhysRevLett.116.251302}, 116, 251302

\bibitem[\protect\citeauthoryear{{Bi{\v{c}}{\'a}k}, {Katz}  \&
  {Lynden-Bell}}{{Bi{\v{c}}{\'a}k} et~al.}{2007}]{Bicak_2007}
{Bi{\v{c}}{\'a}k} J.,  {Katz} J.,   {Lynden-Bell} D.,  2007, \mn@doi [\prd]
  {10.1103/PhysRevD.76.063501}, \href
  {https://ui.adsabs.harvard.edu/abs/2007PhRvD..76f3501B} {76, 063501}

\bibitem[\protect\citeauthoryear{{Bolejko}}{{Bolejko}}{2018a}]{Bolejko2018silent}
{Bolejko} K.,  2018a, \mn@doi [Classical and Quantum Gravity]
  {10.1088/1361-6382/aa9d32}, \href
  {https://ui.adsabs.harvard.edu/abs/2018CQGra..35b4003B} {35, 024003}

\bibitem[\protect\citeauthoryear{{Bolejko}}{{Bolejko}}{2018b}]{Bolejko2018Hubble}
{Bolejko} K.,  2018b, \mn@doi [\prd] {10.1103/PhysRevD.97.103529}, \href
  {https://ui.adsabs.harvard.edu/abs/2018PhRvD..97j3529B} {97, 103529}

\bibitem[\protect\citeauthoryear{{Bolejko} \& {Lasky}}{{Bolejko} \&
  {Lasky}}{2008}]{BolejkoLasky2008}
{Bolejko} K.,  {Lasky} P.~D.,  2008, \mn@doi [\mnras]
  {10.1111/j.1745-3933.2008.00555.x}, \href
  {https://ui.adsabs.harvard.edu/abs/2008MNRAS.391L..59B} {391, L59}

\bibitem[\protect\citeauthoryear{Bolejko, Nazer  \& Wiltshire}{Bolejko
  et~al.}{2016}]{Bolejko2016}
Bolejko K.,  Nazer M.~A.,   Wiltshire D.~L.,  2016, \mn@doi [\jcap]
  {10.1088/1475-7516/2016/06/035}, {\protect\rm06}, 035

\bibitem[\protect\citeauthoryear{{Borzyszkowski}, {Bertacca}  \&
  {Porciani}}{{Borzyszkowski} et~al.}{2017}]{Borzyszkowski2017}
{Borzyszkowski} M.,  {Bertacca} D.,   {Porciani} C.,  2017, \mn@doi [\mnras]
  {10.1093/mnras/stx1423}, \href
  {https://ui.adsabs.harvard.edu/abs/2017MNRAS.471.3899B} {471, 3899}

\bibitem[\protect\citeauthoryear{Bos, van~de Weygaert, Dolag  \& Pettorino}{Bos
  et~al.}{2012}]{Bos2012}
Bos E. G.~P.,  van~de Weygaert R.,  Dolag K.,   Pettorino V.,  2012, \mn@doi
  [\mnras] {10.1111/j.1365-2966.2012.21478.x}, 426, 440

\bibitem[\protect\citeauthoryear{{Brown}, {Diener}, {Sarbach}, {Schnetter}  \&
  {Tiglio}}{{Brown} et~al.}{2009}]{Brown:2009McLachlan}
{Brown} D.,  {Diener} P.,  {Sarbach} O.,  {Schnetter} E.,   {Tiglio} M.,  2009,
  \mn@doi [\prd] {10.1103/PhysRevD.79.044023}, \href
  {https://ui.adsabs.harvard.edu/abs/2009PhRvD..79d4023B} {79, 044023}

\bibitem[\protect\citeauthoryear{{Bruni}, {Thomas}  \& {Wands}}{{Bruni}
  et~al.}{2014}]{Bruni2014}
{Bruni} M.,  {Thomas} D.~B.,   {Wands} D.,  2014, \mn@doi [\prd]
  {10.1103/PhysRevD.89.044010}, \href
  {https://ui.adsabs.harvard.edu/abs/2014PhRvD..89d4010B} {89, 044010}

\bibitem[\protect\citeauthoryear{Buchert}{Buchert}{2000}]{Buchert2000}
Buchert T.,  2000, \mn@doi [Gen. Relativ. Gravitation]
  {10.1023/a:1001800617177}, 32, 105

\bibitem[\protect\citeauthoryear{{Buchert}}{{Buchert}}{2018}]{Buchert2018}
{Buchert} T.,  2018, \mn@doi [\mnras] {10.1093/mnrasl/slx160}, \href
  {https://ui.adsabs.harvard.edu/abs/2018MNRAS.473L..46B} {473, L46}

\bibitem[\protect\citeauthoryear{{Buchert} \& {Ehlers}}{{Buchert} \&
  {Ehlers}}{1997}]{Buchert1997}
{Buchert} T.,  {Ehlers} J.,  1997, \mn@doi [\aap]
  {10.48550/arXiv.astro-ph/9510056}, \href
  {https://ui.adsabs.harvard.edu/abs/1997A&A...320....1B} {320, 1}

\bibitem[\protect\citeauthoryear{{Buchert} \& {R{\"a}s{\"a}nen}}{{Buchert} \&
  {R{\"a}s{\"a}nen}}{2012}]{Buchert:2012A}
{Buchert} T.,  {R{\"a}s{\"a}nen} S.,  2012, \mn@doi [Annu. Rev. of Nuclear and
  Part. Sci.] {10.1146/annurev.nucl.012809.104435}, \href
  {https://ui.adsabs.harvard.edu/abs/2012ARNPS..62...57B} {62, 57}

\bibitem[\protect\citeauthoryear{Buchert, Mourier  \& Roy}{Buchert
  et~al.}{2020}]{Buchert2020}
Buchert T.,  Mourier P.,   Roy X.,  2020, \mn@doi [Gen. Relativ. Gravitation]
  {10.1007/s10714-020-02670-6}, 52

\bibitem[\protect\citeauthoryear{{Campanelli}, {Lousto}, {Marronetti}  \&
  {Zlochower}}{{Campanelli} et~al.}{2006}]{Campanelli:2006}
{Campanelli} M.,  {Lousto} C.~O.,  {Marronetti} P.,   {Zlochower} Y.,  2006,
  \mn@doi [\prl] {10.1103/PhysRevLett.96.111101}, \href
  {https://ui.adsabs.harvard.edu/abs/2006PhRvL..96k1101C} {96, 111101}

\bibitem[\protect\citeauthoryear{Cautun, van~de Weygaert, Jones  \&
  Frenk}{Cautun et~al.}{2014}]{Cautun2014}
Cautun M.,  van~de Weygaert R.,  Jones B. J.~T.,   Frenk C.~S.,  2014, \mn@doi
  [Proc. Int. Astron. Union] {10.1017/s1743921316009613}, 11, 47

\bibitem[\protect\citeauthoryear{Cautun, Cai  \& Frenk}{Cautun
  et~al.}{2016}]{Cautun2016}
Cautun M.,  Cai Y.-C.,   Frenk C.~S.,  2016, \mn@doi [\mnras]
  {10.1093/mnras/stw154}, 457, 2540

\bibitem[\protect\citeauthoryear{{Clifton}, {Gallagher}, {Goldberg}  \&
  {Malik}}{{Clifton} et~al.}{2020}]{Clifton_2020}
{Clifton} T.,  {Gallagher} C.~S.,  {Goldberg} S.,   {Malik} K.~A.,  2020,
  \mn@doi [\prd] {10.1103/PhysRevD.101.063530}, \href
  {https://ui.adsabs.harvard.edu/abs/2020PhRvD.101f3530C} {101, 063530}

\bibitem[\protect\citeauthoryear{{Clough}, {Lim}, {DiNunno}, {Fischler},
  {Flauger}  \& {Paban}}{{Clough} et~al.}{2017}]{Clough:2017}
{Clough} K.,  {Lim} E.~A.,  {DiNunno} B.~S.,  {Fischler} W.,  {Flauger} R.,
  {Paban} S.,  2017, \mn@doi [\jcap] {10.1088/1475-7516/2017/09/025}, \href
  {https://ui.adsabs.harvard.edu/abs/2017JCAP...09..025C} {{\protect\rm09},
  025}

\bibitem[\protect\citeauthoryear{Contarini et~al.,}{Contarini
  et~al.}{2022}]{Contarini2022}
Contarini S.,  et~al., 2022, \mn@doi [\aap] {10.1051/0004-6361/202244095}, 667,
  A162

\bibitem[\protect\citeauthoryear{{Contarini}, {Pisani}, {Hamaus}, {Marulli},
  {Moscardini}  \& {Baldi}}{{Contarini} et~al.}{2023}]{Contarini2023}
{Contarini} S.,  {Pisani} A.,  {Hamaus} N.,  {Marulli} F.,  {Moscardini} L.,
  {Baldi} M.,  2023, \mn@doi [\apj] {10.3847/1538-4357/acde54}, \href
  {https://ui.adsabs.harvard.edu/abs/2023ApJ...953...46C} {953, 46}

\bibitem[\protect\citeauthoryear{Contarini, Pisani, Hamaus, Marulli, Moscardini
   \& Baldi}{Contarini et~al.}{2024}]{Contarini2024}
Contarini S.,  Pisani A.,  Hamaus N.,  Marulli F.,  Moscardini L.,   Baldi M.,
  2024, \mn@doi [\aap] {10.1051/0004-6361/202347572}, 682, A20

\bibitem[\protect\citeauthoryear{Dam}{Dam}{2016}]{Damthesis2016}
Dam L.~H.,  2016, \protect{MSc} thesis, University of Canterbury,
  {Christchurch, NZ}, \mn@doi{10.26021/9152}, \url
  {http://hdl.handle.net/10092/13167}

\bibitem[\protect\citeauthoryear{Daverio, Dirian  \& Mitsou}{Daverio
  et~al.}{2017}]{Daverio2017}
Daverio D.,  Dirian Y.,   Mitsou E.,  2017, \mn@doi [Classical and Quantum
  Gravity] {10.1088/1361-6382/aa9312}, 34, 237001

\bibitem[\protect\citeauthoryear{Daverio, Dirian  \& Mitsou}{Daverio
  et~al.}{2019}]{Daverio2019}
Daverio D.,  Dirian Y.,   Mitsou E.,  2019, \mn@doi [\jcap]
  {10.1088/1475-7516/2019/10/065}, {\protect\rm10}, 065

\bibitem[\protect\citeauthoryear{{Duley}, {Nazer}  \& {Wiltshire}}{{Duley}
  et~al.}{2013}]{DuleyTimescapeRadiation2013}
{Duley} J. A.~G.,  {Nazer} M.~A.,   {Wiltshire} D.~L.,  2013, \mn@doi
  [Classical and Quantum Gravity] {10.1088/0264-9381/30/17/175006}, \href
  {https://ui.adsabs.harvard.edu/abs/2013CQGra..30q5006D} {30, 175006}

\bibitem[\protect\citeauthoryear{East, Wojtak  \& Pretorius}{East
  et~al.}{2019}]{East2019}
East W.~E.,  Wojtak R.,   Pretorius F.,  2019, \mn@doi [\prd]
  {10.1103/physrevd.100.103533}, 100

\bibitem[\protect\citeauthoryear{El‐Ad \& Piran}{El‐Ad \&
  Piran}{1997}]{El_Ad1997}
El‐Ad H.,  Piran T.,  1997, \mn@doi [\apj] {10.1086/304973}, 491, 421

\bibitem[\protect\citeauthoryear{{Fidler}, {Tram}, {Rampf}, {Crittenden},
  {Koyama}  \& {Wands}}{{Fidler} et~al.}{2016}]{Fidler_2016}
{Fidler} C.,  {Tram} T.,  {Rampf} C.,  {Crittenden} R.,  {Koyama} K.,   {Wands}
  D.,  2016, \mn@doi [\jcap] {10.1088/1475-7516/2016/09/031}, \href
  {https://ui.adsabs.harvard.edu/abs/2016JCAP...09..031F} {{\protect\rm09},
  031}

\bibitem[\protect\citeauthoryear{{Fidler}, {Tram}, {Rampf}, {Crittenden},
  {Koyama}  \& {Wands}}{{Fidler} et~al.}{2017}]{Fidler_2017}
{Fidler} C.,  {Tram} T.,  {Rampf} C.,  {Crittenden} R.,  {Koyama} K.,   {Wands}
  D.,  2017, \mn@doi [\jcap] {10.1088/1475-7516/2017/12/022}, \href
  {https://ui.adsabs.harvard.edu/abs/2017JCAP...12..022F} {{\protect\rm12},
  022}

\bibitem[\protect\citeauthoryear{{Giblin} \& {Tishue}}{{Giblin} \&
  {Tishue}}{2019}]{Giblin:2019pre}
{Giblin} J.~T.,  {Tishue} A.~J.,  2019, \mn@doi [\prd]
  {10.1103/PhysRevD.100.063543}, \href
  {https://ui.adsabs.harvard.edu/abs/2019PhRvD.100f3543G} {100, 063543}

\bibitem[\protect\citeauthoryear{Giblin, Mertens  \& Starkman}{Giblin
  et~al.}{2016}]{Giblin2016}
Giblin J.~T.,  Mertens J.~B.,   Starkman G.~D.,  2016, \mn@doi [\prl]
  {10.1103/PhysRevLett.116.251301}, 116, 251301

\bibitem[\protect\citeauthoryear{Giblin, Mertens, Starkman  \& Zentner}{Giblin
  et~al.}{2017}]{Giblin2017}
Giblin J.~T.,  Mertens J.~B.,  Starkman G.~D.,   Zentner A.~R.,  2017, \mn@doi
  [\prd] {10.1103/physrevd.96.103530}, 96

\bibitem[\protect\citeauthoryear{Giblin, Mertens, Starkman  \& Tian}{Giblin
  et~al.}{2019}]{Giblin2019}
Giblin J.~T.,  Mertens J.~B.,  Starkman G.~D.,   Tian C.,  2019, \mn@doi [\prd]
  {10.1103/physrevd.99.023527}, 99, 023527

\bibitem[\protect\citeauthoryear{Granett, Neyrinck  \& Szapudi}{Granett
  et~al.}{2008}]{Granett2008}
Granett B.~R.,  Neyrinck M.~C.,   Szapudi I.,  2008, \mn@doi [\apj]
  {10.1086/591670}, 683, L99

\bibitem[\protect\citeauthoryear{Hamaus, Sutter  \& Wandelt}{Hamaus
  et~al.}{2014}]{Hamaus2014}
Hamaus N.,  Sutter P.~M.,   Wandelt B.~D.,  2014, \mn@doi [\prl]
  {10.1103/physrevlett.112.251302}, 112

\bibitem[\protect\citeauthoryear{{Hamaus}, {Pisani}, {Sutter}, {Lavaux},
  {Escoffier}, {Wandelt}  \& {Weller}}{{Hamaus} et~al.}{2016}]{Hamaus2016}
{Hamaus} N.,  {Pisani} A.,  {Sutter} P.~M.,  {Lavaux} G.,  {Escoffier} S.,
  {Wandelt} B.~D.,   {Weller} J.,  2016, \mn@doi [\prl]
  {10.1103/PhysRevLett.117.091302}, \href
  {https://ui.adsabs.harvard.edu/abs/2016PhRvL.117i1302H} {117, 091302}

\bibitem[\protect\citeauthoryear{{Hamaus}, {Pisani}, {Choi}, {Lavaux},
  {Wandelt}  \& {Weller}}{{Hamaus} et~al.}{2020}]{Hamaus2020}
{Hamaus} N.,  {Pisani} A.,  {Choi} J.-A.,  {Lavaux} G.,  {Wandelt} B.~D.,
  {Weller} J.,  2020, \mn@doi [\jcap] {10.1088/1475-7516/2020/12/023}, \href
  {https://ui.adsabs.harvard.edu/abs/2020JCAP...12..023H} {{\protect\rm12},
  023}

\bibitem[\protect\citeauthoryear{{Heinesen} \& {Macpherson}}{{Heinesen} \&
  {Macpherson}}{2022}]{Heinesen2022}
{Heinesen} A.,  {Macpherson} H.~J.,  2022, \mn@doi [\jcap]
  {10.1088/1475-7516/2022/03/057}, \href
  {https://ui.adsabs.harvard.edu/abs/2022JCAP...03..057H} {{\protect\rm03},
  057}

\bibitem[\protect\citeauthoryear{{Hoyle} \& {Vogeley}}{{Hoyle} \&
  {Vogeley}}{2002}]{Hoyle2002}
{Hoyle} F.,  {Vogeley} M.~S.,  2002, \mn@doi [\apj] {10.1086/338340}, \href
  {https://ui.adsabs.harvard.edu/abs/2002ApJ...566..641H} {566, 641}

\bibitem[\protect\citeauthoryear{{Hoyle} \& {Vogeley}}{{Hoyle} \&
  {Vogeley}}{2004}]{Hoyle2004}
{Hoyle} F.,  {Vogeley} M.~S.,  2004, \mn@doi [\apj] {10.1086/386279}, \href
  {https://ui.adsabs.harvard.edu/abs/2004ApJ...607..751H} {607, 751}

\bibitem[\protect\citeauthoryear{{Ishibashi} \& {Wald}}{{Ishibashi} \&
  {Wald}}{2006}]{Ishibashi2006}
{Ishibashi} A.,  {Wald} R.~M.,  2006, \mn@doi [Classical and Quantum Gravity]
  {10.1088/0264-9381/23/1/012}, \href
  {https://ui.adsabs.harvard.edu/abs/2006CQGra..23..235I} {23, 235}

\bibitem[\protect\citeauthoryear{{Kaiser}}{{Kaiser}}{2017}]{Kaiser2017}
{Kaiser} N.,  2017, \mn@doi [\mnras] {10.1093/mnras/stx907}, \href
  {https://ui.adsabs.harvard.edu/abs/2017MNRAS.469..744K} {469, 744}

\bibitem[\protect\citeauthoryear{{Kov{\'a}cs}, {Beck}, {Szapudi}, {Csabai},
  {R{\'a}cz}  \& {Dobos}}{{Kov{\'a}cs} et~al.}{2020}]{Kovacs_2020}
{Kov{\'a}cs} A.,  {Beck} R.,  {Szapudi} I.,  {Csabai} I.,  {R{\'a}cz} G.,
  {Dobos} L.,  2020, \mn@doi [\mnras] {10.1093/mnras/staa2631}, \href
  {https://ui.adsabs.harvard.edu/abs/2020MNRAS.499..320K} {499, 320}

\bibitem[\protect\citeauthoryear{{Kov{\'a}cs}, {Beck}, {Smith}, {R{\'a}cz},
  {Csabai}  \& {Szapudi}}{{Kov{\'a}cs} et~al.}{2022}]{Kovacs_2022}
{Kov{\'a}cs} A.,  {Beck} R.,  {Smith} A.,  {R{\'a}cz} G.,  {Csabai} I.,
  {Szapudi} I.,  2022, \mn@doi [\mnras] {10.1093/mnras/stac903}, \href
  {https://ui.adsabs.harvard.edu/abs/2022MNRAS.513...15K} {513, 15}

\bibitem[\protect\citeauthoryear{{Lahav}, {Lilje}, {Primack}  \&
  {Rees}}{{Lahav} et~al.}{1991}]{Lahav1991}
{Lahav} O.,  {Lilje} P.~B.,  {Primack} J.~R.,   {Rees} M.~J.,  1991, \mn@doi
  [\mnras] {10.1093/mnras/251.1.128}, \href
  {https://ui.adsabs.harvard.edu/abs/1991MNRAS.251..128L} {251, 128}

\bibitem[\protect\citeauthoryear{Lavaux \& Wandelt}{Lavaux \&
  Wandelt}{2010}]{Lavaux2010}
Lavaux G.,  Wandelt B.~D.,  2010, \mn@doi [\mnras]
  {10.1111/j.1365-2966.2010.16197.x}, 403, 1392

\bibitem[\protect\citeauthoryear{{Lehner} \& {Pretorius}}{{Lehner} \&
  {Pretorius}}{2014}]{Lehner:2014}
{Lehner} L.,  {Pretorius} F.,  2014, \mn@doi [\araa]
  {10.1146/annurev-astro-081913-040031}, \href
  {https://ui.adsabs.harvard.edu/abs/2014ARA&A..52..661L} {52, 661}

\bibitem[\protect\citeauthoryear{{Lesgourgues}}{{Lesgourgues}}{2011}]{CLASS}
{Lesgourgues} J.,  2011, \mn@doi [arXiv e-prints] {10.48550/arXiv.1104.2932},
  \href {https://ui.adsabs.harvard.edu/abs/2011arXiv1104.2932L} {p.
  arXiv:1104.2932}

\bibitem[\protect\citeauthoryear{Macpherson}{Macpherson}{2023}]{Macpherson2023}
Macpherson H.~J.,  2023, \mn@doi [\jcap] {10.1088/1475-7516/2023/03/019},
  {\protect\rm03}, 019

\bibitem[\protect\citeauthoryear{Macpherson, Lasky  \& Price}{Macpherson
  et~al.}{2017}]{Macpherson2017}
Macpherson H.~J.,  Lasky P.~D.,   Price D.~J.,  2017, \mn@doi [\prd]
  {10.1103/physrevd.95.064028}, 95, 064028

\bibitem[\protect\citeauthoryear{Macpherson, Lasky  \& Price}{Macpherson
  et~al.}{2018}]{Macpherson2018}
Macpherson H.~J.,  Lasky P.~D.,   Price D.~J.,  2018, \mn@doi [\apj]
  {10.3847/2041-8213/aadf8c}, 865, L4

\bibitem[\protect\citeauthoryear{Macpherson, Price  \& Lasky}{Macpherson
  et~al.}{2019}]{Macpherson2019}
Macpherson H.~J.,  Price D.~J.,   Lasky P.~D.,  2019, \mn@doi [\prd]
  {10.1103/PhysRevD.99.063522}, 99, 063522

\bibitem[\protect\citeauthoryear{{Magnall}, {Price}, {Lasky}  \&
  {Macpherson}}{{Magnall} et~al.}{2023}]{Magnall2023}
{Magnall} S.~J.,  {Price} D.~J.,  {Lasky} P.~D.,   {Macpherson} H.~J.,  2023,
  \mn@doi [\prd] {10.1103/PhysRevD.108.103534}, \href
  {https://ui.adsabs.harvard.edu/abs/2023PhRvD.108j3534M} {108, 103534}

\bibitem[\protect\citeauthoryear{{McKay} \& {Wiltshire}}{{McKay} \&
  {Wiltshire}}{2016}]{McKay2016}
{McKay} J.~H.,  {Wiltshire} D.~L.,  2016, \mn@doi [\mnras]
  {10.1093/mnras/stw128}, \href
  {https://ui.adsabs.harvard.edu/abs/2016MNRAS.457.3285M} {457, 3285}

\bibitem[\protect\citeauthoryear{Mertens, Giblin  \& Starkman}{Mertens
  et~al.}{2016}]{Mertens2016}
Mertens J.~B.,  Giblin J.~T.,   Starkman G.~D.,  2016, \mn@doi [\prd]
  {10.1103/PhysRevD.93.124059}, 93, 124059

\bibitem[\protect\citeauthoryear{{Milillo}, {Bertacca}, {Bruni}  \&
  {Maselli}}{{Milillo} et~al.}{2015}]{Milillo2015}
{Milillo} I.,  {Bertacca} D.,  {Bruni} M.,   {Maselli} A.,  2015, \mn@doi
  [\prd] {10.1103/PhysRevD.92.023519}, \href
  {https://ui.adsabs.harvard.edu/abs/2015PhRvD..92b3519M} {92, 023519}

\bibitem[\protect\citeauthoryear{Moresco et~al.,}{Moresco
  et~al.}{2022}]{Moresco2022}
Moresco M.,  et~al., 2022, \mn@doi [Living Rev. Relativ.]
  {10.1007/s41114-022-00040-z}, 25

\bibitem[\protect\citeauthoryear{Nadathur \& Hotchkiss}{Nadathur \&
  Hotchkiss}{2014}]{Nadathur2014}
Nadathur S.,  Hotchkiss S.,  2014, \mn@doi [\mnras] {10.1093/mnras/stu349},
  440, 1248

\bibitem[\protect\citeauthoryear{Nadathur \& Hotchkiss}{Nadathur \&
  Hotchkiss}{2015}]{Nadathur2015}
Nadathur S.,  Hotchkiss S.,  2015, \mn@doi [\mnras] {10.1093/mnras/stv2131},
  454, 2228

\bibitem[\protect\citeauthoryear{Nakamura, Oohara  \& Kojima}{Nakamura
  et~al.}{1987}]{Nakamura1987}
Nakamura T.,  Oohara K.,   Kojima Y.,  1987, \mn@doi [Progress of Theor. Phys.
  Suppl.] {10.1143/PTPS.90.1}, 90, 1

\bibitem[\protect\citeauthoryear{Neyrinck}{Neyrinck}{2008}]{Neyrinck2008}
Neyrinck M.~C.,  2008, \mn@doi [\mnras] {10.1111/j.1365-2966.2008.13180.x},
  386, 2101

\bibitem[\protect\citeauthoryear{Pan, Vogeley, Hoyle, Choi  \& Park}{Pan
  et~al.}{2012}]{Pan_2012}
Pan D.~C.,  Vogeley M.~S.,  Hoyle F.,  Choi Y.-Y.,   Park C.,  2012, \mn@doi
  [\mnras] {10.1111/j.1365-2966.2011.20197.x}, 421, 926

\bibitem[\protect\citeauthoryear{{Peebles}}{{Peebles}}{1993}]{Peebles_1993}
{Peebles} P.~J.~E.,  1993, {Principles of Physical Cosmology}.
Princeton University Press, \mn@doi{10.1515/9780691206721}

\bibitem[\protect\citeauthoryear{{Peebles}}{{Peebles}}{2001}]{Peebles_2001}
{Peebles} P.~J.~E.,  2001, \mn@doi [\apj] {10.1086/322254}, \href
  {https://ui.adsabs.harvard.edu/abs/2001ApJ...557..495P} {557, 495}

\bibitem[\protect\citeauthoryear{Platen, {\VAN{Weygaert}{van de}{van
  de}}~Weygaert  \& Jones}{Platen et~al.}{2007}]{Platen_2007}
Platen E.,  {\VAN{Weygaert}{van de}{van de}}~Weygaert R.,   Jones B. J.~T.,
  2007, \mn@doi [\mnras] {10.1111/j.1365-2966.2007.12125.x}, 380, 551

\bibitem[\protect\citeauthoryear{Pretorius}{Pretorius}{2005}]{Pretorius2005}
Pretorius F.,  2005, \mn@doi [\prl] {10.1103/physrevlett.95.121101}, 95, 121101

\bibitem[\protect\citeauthoryear{Rácz, Dobos, Beck, Szapudi  \& Csabai}{Rácz
  et~al.}{2017}]{Racz2017}
Rácz G.,  Dobos L.,  Beck R.,  Szapudi I.,   Csabai I.,  2017, \mn@doi
  [\mnras] {10.1093/mnrasl/slx026}, 469, L1

\bibitem[\protect\citeauthoryear{Schuster, Hamaus, Dolag  \& Weller}{Schuster
  et~al.}{2023}]{Schuster2023}
Schuster N.,  Hamaus N.,  Dolag K.,   Weller J.,  2023, \mn@doi [\jcap]
  {10.1088/1475-7516/2023/05/031}, {\protect\rm05}, 031

\bibitem[\protect\citeauthoryear{Sheth \& {\VAN{Weygaert}{van de}{van
  de}}~Weygaert}{Sheth \& {\VAN{Weygaert}{van de}{van
  de}}~Weygaert}{2004}]{Sheth_2004}
Sheth R.~K.,  {\VAN{Weygaert}{van de}{van de}}~Weygaert R.,  2004, \mn@doi
  [\mnras] {10.1111/j.1365-2966.2004.07661.x}, 350, 517

\bibitem[\protect\citeauthoryear{Shibata \& Nakamura}{Shibata \&
  Nakamura}{1995}]{Shibata1995}
Shibata M.,  Nakamura T.,  1995, \mn@doi [\prd] {10.1103/PhysRevD.52.5428}, 52,
  5428

\bibitem[\protect\citeauthoryear{{Shibata}, {Taniguchi}  \&
  {Ury{\={u}}}}{{Shibata} et~al.}{2005}]{Shibata:2005}
{Shibata} M.,  {Taniguchi} K.,   {Ury{\={u}}} K.,  2005, \mn@doi [\prd]
  {10.1103/PhysRevD.71.084021}, \href
  {https://ui.adsabs.harvard.edu/abs/2005PhRvD..71h4021S} {71, 084021}

\bibitem[\protect\citeauthoryear{Stopyra, Peiris  \& Pontzen}{Stopyra
  et~al.}{2020}]{Stopyra2020}
Stopyra S.,  Peiris H.~V.,   Pontzen A.,  2020, \mn@doi [\mnras]
  {10.1093/mnras/staa3587}, 500, 4173

\bibitem[\protect\citeauthoryear{Sutter, Lavaux, Wandelt  \& Weinberg}{Sutter
  et~al.}{2012}]{Sutter2012}
Sutter P.~M.,  Lavaux G.,  Wandelt B.~D.,   Weinberg D.~H.,  2012, \mn@doi
  [\apj] {10.1088/0004-637x/761/1/44}, 761, 44

\bibitem[\protect\citeauthoryear{Sutter, Pisani, Wandelt  \& Weinberg}{Sutter
  et~al.}{2014a}]{Sutter2014AP}
Sutter P.~M.,  Pisani A.,  Wandelt B.~D.,   Weinberg D.~H.,  2014a, \mn@doi
  [\mnras] {10.1093/mnras/stu1392}, 443, 2983

\bibitem[\protect\citeauthoryear{Sutter, Elahi, Falck, Onions, Hamaus, Knebe,
  Srisawat  \& Schneider}{Sutter et~al.}{2014b}]{Sutter2014VoidLife}
Sutter P.~M.,  Elahi P.,  Falck B.,  Onions J.,  Hamaus N.,  Knebe A.,
  Srisawat C.,   Schneider A.,  2014b, \mn@doi [\mnras]
  {10.1093/mnras/stu1845}, 445, 1235

\bibitem[\protect\citeauthoryear{{Sutter} et~al.,}{{Sutter}
  et~al.}{2015}]{sutter2014vide}
{Sutter} P.~M.,  et~al., 2015, \mn@doi [Astronomy and Computing]
  {10.1016/j.ascom.2014.10.002}, \href
  {https://ui.adsabs.harvard.edu/abs/2015A&C.....9....1S} {9, 1}

\bibitem[\protect\citeauthoryear{{Thomas}, {Bruni}  \& {Wands}}{{Thomas}
  et~al.}{2015a}]{Thomas2015b}
{Thomas} D.~B.,  {Bruni} M.,   {Wands} D.,  2015a, \mn@doi [\jcap]
  {10.1088/1475-7516/2015/09/021}, \href
  {https://ui.adsabs.harvard.edu/abs/2015JCAP...09..021T} {{\protect\rm09},
  021}

\bibitem[\protect\citeauthoryear{{Thomas}, {Bruni}  \& {Wands}}{{Thomas}
  et~al.}{2015b}]{Thomas2015}
{Thomas} D.~B.,  {Bruni} M.,   {Wands} D.,  2015b, \mn@doi [\mnras]
  {10.1093/mnras/stv1390}, \href
  {https://ui.adsabs.harvard.edu/abs/2015MNRAS.452.1727T} {452, 1727}

\bibitem[\protect\citeauthoryear{{Tian}, {Anselmi}, {Carney}, {Giblin},
  {Mertens}  \& {Starkman}}{{Tian} et~al.}{2021}]{Tian:2021}
{Tian} C.,  {Anselmi} S.,  {Carney} M.~F.,  {Giblin} J.~T.,  {Mertens} J.,
  {Starkman} G.,  2021, \mn@doi [\prd] {10.1103/PhysRevD.103.083513}, \href
  {https://ui.adsabs.harvard.edu/abs/2021PhRvD.103h3513T} {103, 083513}

\bibitem[\protect\citeauthoryear{Tikhonov \& Karachentsev}{Tikhonov \&
  Karachentsev}{2006}]{Tikhonov2006}
Tikhonov A.~V.,  Karachentsev I.~D.,  2006, \mn@doi [\apj] {10.1086/508981},
  653, 969

\bibitem[\protect\citeauthoryear{{Tram}, {Brandbyge}, {Dakin}  \&
  {Hannestad}}{{Tram} et~al.}{2019}]{Tram2019}
{Tram} T.,  {Brandbyge} J.,  {Dakin} J.,   {Hannestad} S.,  2019, \mn@doi
  [\jcap] {10.1088/1475-7516/2019/03/022}, \href
  {https://ui.adsabs.harvard.edu/abs/2019JCAP...03..022T} {{\protect\rm03},
  022}

\bibitem[\protect\citeauthoryear{Wang}{Wang}{2018}]{Wang2018}
Wang K.,  2018, \mn@doi [European Phys. J. C] {10.1140/epjc/s10052-018-6103-7},
  78

\bibitem[\protect\citeauthoryear{Werneck et~al.,}{Werneck
  et~al.}{2023}]{EinsteinToolkit:2023_05}
Werneck L.,  et~al., 2023, The Einstein Toolkit,
  \mn@doi{10.5281/zenodo.7942541}

\bibitem[\protect\citeauthoryear{{Wiltshire}}{{Wiltshire}}{2007a}]{wiltshireCosmicClocks2007a}
{Wiltshire} D.~L.,  2007a, \mn@doi [New J. of Phys.]
  {10.1088/1367-2630/9/10/377}, \href
  {https://ui.adsabs.harvard.edu/abs/2007NJPh....9..377W} {9, 377}

\bibitem[\protect\citeauthoryear{{Wiltshire}}{{Wiltshire}}{2007b}]{wiltshireExactSolution2007b}
{Wiltshire} D.~L.,  2007b, \mn@doi [\prl] {10.1103/PhysRevLett.99.251101},
  \href {https://ui.adsabs.harvard.edu/abs/2007PhRvL..99y1101W} {99, 251101}

\bibitem[\protect\citeauthoryear{Wiltshire}{Wiltshire}{2009}]{wiltshireAverageObservationalQuantities2009}
Wiltshire D.~L.,  2009, \mn@doi [\prd] {10.1103/PhysRevD.80.123512}, 80, 123512

\makeatother
\end{thebibliography}


\appendix

\section{Resolution study}
\label{a:resolution}
While our two simulations have different physical resolutions, they have the same numerical resolution of $256^3$ grid cells.
This value, not physical resolution, is what affects numerical error. 
To determine whether numerical error is having a significant effect on the void statistics,
we analyse a second simulation with \perhMpc{4} physical resolution
which instead has only $128^3$ grid cells, 
so the simulation volume is \perhMpc{512} across.
The lower numerical resolution means this simulation has greater numerical error.
If numerical error is significantly contributing to the statistics, they will differ between the two \perhMpc{4} resolution simulations.

\begin{figure}
	\includegraphics{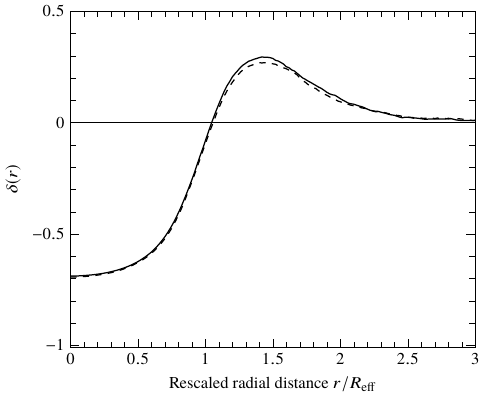}
    \caption{
    Comparison of mean density contrast on spherical shells in a stack of 2000 voids from a simulation with $128^3$ cells (solid line)
    and 2000 voids from the main \perhMpc{4} resolution simulation with $256^3$ cells (dashed line).
    }
    \label{fig:resolution_study_density}
\end{figure}

\begin{figure}
	\includegraphics{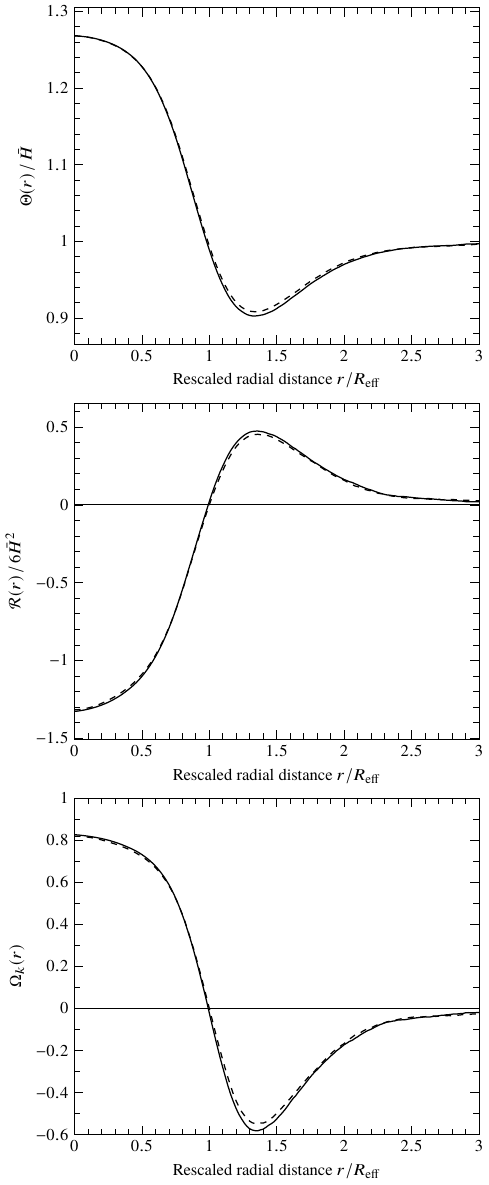}
    \caption{
    Comparison of $\Theta$, $\threericci$, $\OK$ (top to bottom) on spherical shells in a stack of 2000 voids from a simulation with $128^3$ cells (solid lines)
    and 2000 voids from the main \perhMpc{4} resolution simulation with $256^3$ cells (dashed lines).
    }
    \label{fig:resolution_study_curvature}
\end{figure}

Fig.~\ref{fig:resolution_study_density} shows the stacked mean radial density contrast for 2000 randomly-selected voids in the $256^3$ cell, \perhMpc{4} resolution simulation (dashed curve)
along with the equivalent for the $128^3$ simulation (solid curve).
Fig.~\ref{fig:resolution_study_curvature} does the same for the three other scalars, $\Theta$, $\threericci$, and $\OK$, from Figs.~\ref{fig:radial_theta}, \ref{fig:radial_ricci}, and \ref{fig:radial_omega} respectively.
The closeness of the two curves in each of these figures shows that numerical error is not contributing significantly.

The 2000 voids picked from the $128^3$ simulation are, by necessity, different from the 2000 picked out of the $256^3$ simulation.
Thus, this comparison also shows that the statistics are not significantly affected by choosing one random set of 2000 voids from the population over another random set.



\bsp	
\label{lastpage}
\end{document}